\documentclass[journal]{IEEEtran}
\usepackage{cite}
\usepackage{amsmath,amssymb,amsfonts,amsthm,mathrsfs}
\usepackage{bm}
\usepackage{graphicx}
\usepackage{multirow}
\usepackage{subfigure}
\usepackage{textcomp}
\usepackage{enumerate,paralist}
\usepackage{mathrsfs} 
\usepackage{setspace}
\usepackage{nomencl}
\usepackage{algpseudocode,algorithm}
\usepackage{threeparttable}
\usepackage{comment}

\makenomenclature

\theoremstyle{definition}
\newtheorem{theorem}{\bf Theorem}

\newtheorem{remark}{\bf Remark}

\newtheorem{definition}{Definition}

\graphicspath{{Graphics/}}	
\usepackage[colorlinks, linkcolor=black, anchorcolor=black, citecolor=black, urlcolor =black]{hyperref}
\newcommand{\norm}[1]{\left\lVert#1\right\rVert}

\begin{document}
\title{Multi-kernel Correntropy-based Orientation Estimation of IMUs: Gradient Descent Methods}
\author{Shilei Li, Lijing Li, Dawei Shi, Yunjiang Lou, Ling Shi
\thanks{Shilei Li and Ling Shi are with the Department of Electronic and Computer Engineering, The Hong Kong University of Science and Technology, Hong Kong, China (e-mail: slidk@connect.ust.hk, eesling@ust.hk).}
\thanks{Lijing Li is with the School of Information and Control Engineering, China University of Mining and Technology, China (e-mail: lilijing\_29@163.com).}
\thanks{Dawei Shi is with the School of Automation, Beijing Institute of Technology, China (e-mail: daweishi@bit.edu.cn).}
\thanks{Yunjiang Lou is with the State Key Laboratory of Robotics and System, School of Mechanical Engineering and Automation, Harbin Institute of Technology Shenzhen, Shenzhen 518055, China (e-mail: louyj@hit.edu.cn).}
}
\maketitle

\begin{abstract}
This paper presents two computationally efficient algorithms for the orientation estimation of inertial measurement units (IMUs): the multi-kernel correntropy-based gradient descent (CGD) and the multi-kernel correntropy-based decoupled orientation estimation (CDOE). Traditional methods, such as gradient descent (GD) and decoupled orientation estimation (DOE), rely on the least square (LS) criterion in algorithm derivation, making them vulnerable to external acceleration and magnetic interference. To address this issue, we first demonstrate that the multi-kernel correntropy loss (MKCL) is an optimal objective function under the maximum likelihood estimation (MLE) framework when the noise follows a  specific type of heavy-tailed distribution. \textcolor{black}{Then, we provide some important properties of the MKCL as a cost function.} By replacing the LS cost with the MKCL, we develop the CGD and CDOE algorithms. We evaluate the effectiveness of our proposed methods by comparing them with existing algorithms in various situations. Experimental results indicate that our proposed methods (CGD and CDOE) outperform their conventional counterparts (GD and DOE), especially when faced with external acceleration and magnetic disturbances. Furthermore, the new algorithms demonstrate significantly lower computational complexity than Kalman filter-based approaches, making them suitable for applications with low-cost microprocessors. 
\end{abstract}
\begin{IEEEkeywords}
	multi-kernel correntropy,  gradient descent, orientation estimation, IMUs
\end{IEEEkeywords}

\section{Introduction}
The use of micro-electro-mechanical-based IMUs has become widespread in various fields, including robotics~\cite{a1}, navigation~\cite{a2}, and localization~\cite{az3}. Additionally, they can be found in many consumer electronics, such as smartphones, tablet computers, and smartwatches. \textcolor{black}{IMUs can provide information about the carrier's orientation and can give additional localization information when combined with other sensors (e.g., ultra-wideband, camera, ultrasound, and GPS~\cite{c15,e5}). Compared with optical motion capture systems that suffer from non-line-of-sight conditions, limited capture range, expensiveness, and post-processing, IMUs have the advantages of low cost, small size, portability, and real-time processing}. Recently, IMUs have been utilized in health monitoring~\cite{az4}, joint angle estimation~\cite{az5}, and robot control~\cite{az7}. These applications require IMUs to maintain specific precision in a disturbed environment and to execute at a faster sampling frequency. Therefore, the development of computationally efficient and robust algorithms is necessary.

IMUs are composed of gyroscopes, accelerometers, and magnetometers. The gyroscope measures the angular rate and the orientation can be obtained by integrating gyroscope readings. However, long-term integration may cause an unbounded orientation error due to gyroscope drift and numerical integration errors. To this end, noisy but drift-free accelerometers and magnetometers are utilized to correct the orientation estimation. The orientation estimation algorithms of IMUs can be broadly categorized into three groups: the complementary filter, the Kalman filter (KF), and the gradient descent (GD). The complementary filter was initially proposed by Mahony et al. ~\cite{b1,b2} for aerospace applications, which was designed on the special orthogonal group ($SO_3$) and offered robustness to noise. The KF-based methods, which are widely used in commercial IMUs, have higher accuracy than complementary filters but require higher computational complexity and more challenging parameter-tuning procedures. Typical algorithms of this type include the extended KF~\cite{b3,bz3}, the indirect KF~\cite{b4}, and the error state KF (ESKF)~\cite{b5,b6,c10}. The GDs, proposed by Madgwick et al.~\cite{b7}, and its enhanced version by Seel et al.~\cite{b14}, have fewer parameters to tune and are reported of owing similar performance with the Kalman filter-based methods. Note that some algorithms may fall outside of the aforementioned three categories, such as the sliding mode observer~\cite{b8} and the Luenberger observer~\cite{b9}. In this work, we use the GDs~\cite{b7,b14} and ESKF~\cite{b6,c10} as the benchmark methods due to their popularity.

The challenges associated with IMU algorithms are their heavy computation cost and sensitivity to external disturbances. The high computation complexity would induce a low execution frequency, which may hinder the real-time control of robots~\cite{az4}. Another challenge is the accuracy degeneration due to external disturbances, i.e., external acceleration and magnetic interference. Many strategies have been deployed to mitigate the negative impacts of these disturbances. One of the most straightforward approaches may be the normalization of the accelerometer and magnetometer readings, which was employed in ~\cite{b2,b7,b14}. Another way is to model the disturbance as a first-order Markov model and augment it as a new state to attenuate its bad effects~\cite{b3,b4,b5,b6}. Some other strategies, e.g., covariance matrices adaption~\cite{b4,b6}, disturbance detection~\cite{c19}, and the combination of them, are also frequently used to increase the robustness of the algorithm. In some indoor applications, magnetic disturbance, caused by surrounding ferromagnetic materials (e.g., iron, steel, and magnets), is a major reason for deteriorating both the \textcolor{black}{heading} (yaw) and the inclination (roll and pitch) accuracy. To handle this issue, Seel et al. proposed a decoupled orientation estimation (DOE) scheme that designed an analytical solution for sensor fusion of IMUs so that the magnetometer readings only affected the heading estimation~\cite{b14}. A similar idea is also deployed in~\cite{c20} where a decoupled quaternion solution was developed.  

\textcolor{black}{Although many algorithms have been developed to counteract the defects caused by disturbance, they directly or indirectly relied on the Gaussian noise assumption and utilized the LS criterion in algorithm derivation~\cite{b15,b16,b6,b7,b14}. In this work, we demonstrate that great freedom can be obtained by generalizing the LS to the \emph{multi-kernel correntropy loss} (MKCL) where the underlying MLE interpretation is to extend the Gaussian assumption to a type of heavy-tailed distribution. The correntropy is the local similarity measurement of two random variables. It has been successfully utilized in KF~\cite{c2,c3,c13}, adaptive filtering~\cite{c5}, and machine learning~\cite{c21}. In \cite{c22}, we extended the definition of correntropy from random variables to random vectors and defined the terminology \emph{multi-kernel correntropy} (MKC) which greatly alleviates the conservatism of the traditional correntropy (note that the MKC proposed in our previous work is different with the concept in ~\cite{c21}).} Based on this idea, in \cite{b15,b16}, we proposed two MKC-based KFs for orientation estimation of IMUs, which achieved satisfactory performance both with and without disturbance~\cite{b15,b16}.

Although the performances of ~\cite{b15,b16} are satisfactory compared with ~\cite{b6,b7,b14}, they have some drawbacks. Firstly, they require accurate covariance matrices of sensors, which increases the difficulty of implementation. Secondly, these algorithms possess heavy computational complexity and are not suitable for low-cost IMUs with limited computational capability. Thirdly, the relationship between the objective function and the underlying noise distribution is not very clear. In this paper, we derive two computationally efficient algorithms, i.e., the CGD and CDOE, which are built upon the GD~\cite{b7} and DOE ~\cite{b14} and utilize the MKCL as objective functions. \textcolor{black}{Specifically, we first demonstrate that accelerometer and magnetometer noises generally are heavy-tailed the Gaussian assumption is not suitable. Then, we provide some important properties of the MKCL and reveal that the MKCL is linked to a type of heavy-tailed distribution. Finally, we derive two novel algorithms by substituting the LS criterion with the MKCL. It is worth mentioning that the aim of this work is not to develop the ``best" orientation estimation algorithms but to illustrate the performance of many existing algorithms can be further improved by matching the objective function with the underlying noise distribution.} The contributions of this paper are summarized as follows.
\color{black}
\begin{enumerate}[1)]
\item  Based on the maximum likelihood estimation (MLE) framework, we demonstrate that the MKCL is an optimal objective function when the noise follows a type of heavy-tailed distribution. Moreover, some important properties of the MKCL are given. 
\item We replace the LS with the MKCL and develop two new algorithms, namely the CGD and CDOE. The newly derived algorithms are robust to external acceleration and magnetic disturbance. Moreover, the CDOE retains the advantage that the magnetometer readings do not affect the inclination estimation.
\item We conduct intensive and comprehensive experiments to verify the performance of the proposed algorithms. The results demonstrate that the two proposed methods outperform their traditional counterparts (i.e., GD and DOE in~\cite{b7,b14}), especially when dealing with external disturbances. Additionally, the two proposed algorithms have comparable accuracy with our previous correntropy-based KF approach~\cite{b16}, but with significantly lower computational cost. 
\end{enumerate}
\color{black}
The remainder of this paper is organized as follows. In Section II, we give sensor models of IMUs, provide some properties of the MKCL, and present a general algorithm framework for IMUs. In Section III, we derive two correntropy-based algorithms (i.e., the CGD and CDOE). In Section IV, we validate the performance of the proposed algorithms. In Section V, we draw a conclusion.
\section{Preliminaries and Problem Formulation}
We first provide the sensor models of IMUs and give the properties of the MKCL. Then, we introduce a general framework for sensor fusion of IMUs and reveal that existing algorithms largely rely on the LS criterion. Finally, we demonstrate that the robustness of many algorithms can be enhanced by substituting the LS with the MKCL.
\subsection{Multi-kernel Correntropy}
The correntropy is a local similarity measure of two random variables $X,Y \in \mathbb{R}$ with 
\begin{equation}
V(X,Y)= E[\kappa(X,Y)]=\int \kappa(x,y) d F_{XY}(x,y)
\end{equation}
where $E[\cdot]$ is the expectation operator, $\kappa(x,y)$ is a shift-invariant Mercer kernel, $F_{XY}(x,y)$ is the joint distribution, and $x$ and $y$ are realizations of $X$ and $Y$. The kernel utilized in this paper is the squared exponential function, i.e., $\kappa(x,y)=G_{\sigma}(x,y)=\exp(-\frac{e^2}{2\sigma^2})$
where $\sigma$ is the kernel bandwidth and $e=x-y$ is the realization error. In ~\cite{b16}, we extended the correntropy from random variables to random vectors and proposed the \emph{multi-kernel correntropy} (MKC) for random vectors $\mathcal{X}, \mathcal{Y} \in \mathbb{R}^{l}$: 
\begin{equation}
V(\mathcal{X},\mathcal{Y})= \sum_{i=1}^{l} E[\sigma_i^2 \kappa_i(\mathcal{X}_i,\mathcal{Y}_i)]
\end{equation}
with 
\begin{equation}\nonumber
\begin{aligned}
E[\sigma_i^2\kappa_i(\mathcal{X}_i,\mathcal{Y}_i)]&=\int \sigma_i^2\kappa_i(x_i,y_i)d F_{\mathcal{X}_i\mathcal{Y}_i}(x_i,y_i)\\
\kappa_i(x_i,y_i)&=G_{\sigma_i}(x_i,y_i)=\exp(-\frac{e_i^2}{2\sigma_i^2})
\end{aligned}
\end{equation}
where $\sigma_i$ is the bandwidth for random pair $\mathcal{X}_i$ and $\mathcal{Y}_i$, $e_i=x_i-y_i$ is the realization error for channel $i$, and $F_{\mathcal{X}_i\mathcal{Y}_i}(x_i,y_i)$ is the joint distribution. In many applications, $F_{\mathcal{X}_i\mathcal{Y}_i}(x_i,y_i)$ is not available whereas only samples $x_{i,k}$ and $y_{i,k}$ can be obtained. In this case, the simple mean estimator is utilized and MKC can be estimated as
 \begin{equation}
 \hat{V}(\mathcal{X},\mathcal{Y})=\sum_{i=1}^{l}\sigma_i^2 \hat{V}_{i}(\mathcal{X}_i,\mathcal{Y}_i)
 \end{equation}
  with
  $\hat{V}_{i}(\mathcal{X}_i,\mathcal{Y}_i)= \frac{1}{N}\sum_{k=1}^{N}G_{\sigma_i}\big({x_{i,k},y_{i,k}}\big).$
  Then, the MKCL can be written as 
  \begin{equation}
  \begin{aligned}
  J_{CL}(e_k) &=\sum_{i=1}^{l}\sigma_i^2 (1-\hat{V}_{i})=\frac{1}{N}\sum_{k=1}^{N}\sum_{i=1}^{l}\sigma_i^2 \Big(1-G_{\sigma_i}(e_{i,k})\Big).
  \label{GL}
  \end{aligned}
  \end{equation}
  where $e_{i,k}=x_{i,k}-y_{i,k}$ and $e_k=[e_{1,k},e_{2,k},\ldots,e_{l,k}]^{T}$. Correspondingly, the LS cost can be expressed as
  \begin{equation}
  \begin{aligned}
  J_{LS}(e_k)& = \frac{1}{2N} \sum_{k=1}^{N}e_k^{T}e_k = \frac{1}{2N}\sum_{k=1}^{N}\sum_{i=1}^{l}e_{i,k}^2.
  \label{LS}
  \end{aligned}
  \end{equation}
  \begin{theorem}
  $J_{CL}$ and $J_{LS}$ in \eqref{GL} and \eqref{LS} are identical when setting $\sigma_i \to \infty$ for $i=1,2,\ldots,l$.
  \label{theorem1}
  \end{theorem}
  The proof of this theorem is shown in Appendix \ref{proof1}.
  \begin{remark}
  	\textcolor{black}{It is worth mentioning that the MKC proposed in our previous work ~\cite{b16} is different from the conventional correntropy in two folds: it utilizes different kernel bandwidths for different random pair of variables. Moreover, specified weights are associated with the MKC so that the MKCL is compatible with the LS criterion.} 
  \end{remark}
  \color{black}
  \subsection{\textcolor{black}{Sensor Models and Sensor Fusion Frameworks}}
  \label{sensormodel}
  In this section, we first introduce the sensor models of IMUs. Then, we provide general frameworks for orientation estimation. Finally, we explain the conventional Gaussian assumption for accelerometers and magnetometers is not suitable.
  	\begin{definition}
  	The orientation from frame $A$ to frame $B$ can be represented by a unit quaternion $^{B}_{A}q \in \mathbb{R}^{4}$ with
  	\begin{equation}\nonumber
  		^{B}_{A}q=\begin{bmatrix}
  			\cos(\alpha/2)\\
  			\sin(\alpha/2)x_{rot}
  		\end{bmatrix}, \|^{B}_{A}q\|_2=1
  	\end{equation}
  	where $\alpha \in \mathbb{R}$ is the rotation angle and $x_{rot} \in \mathbb{R}^{3}$ is the rotation axis with $\|x_{rot}\|_2=1$. In the following section, we restrain the quaternion to be a unit quaternion when representing the orientation without further mentioning. Moreover, the multiplication of two quaternions $p$ and $q$ is denoted as $p \otimes q$.
  \end{definition} 
  The gyroscope model is given by 
  \begin{equation}
  	\begin{aligned}
  		\mathrm{y}_{G,k}&=\mathrm{w}_k+\mathrm{b}_{k}+\mathrm{v}_{G,k}
  		\label{gyr}
  	\end{aligned}
  \end{equation}
  where  $\mathrm{y}_{G,k} \in {\mathbb{R}^{3}}$ is the sensor reading at time step $k$, $\mathrm{w}_k$ is the  angular velocity, $\mathrm{b}_{k}$ is a slow-varying bias, and $\mathrm{v}_{G,k}$ is the noise. The accelerometer model has
  \begin{equation}
  	\begin{aligned}
  		\mathrm{y}_{A,k}&=-{}^{S}\mathrm{g}_{k}+{}^{S}\mathrm{a}_{k}+\mathrm{v}_{A,k}\\
  		{}^{S}\mathrm{g}_{k}&=R({}_{E}^{S}q_k){}^{E}\mathrm{g}
  		\label{accm}
  	\end{aligned}
  \end{equation}
  where $\mathrm{y}_{A,k} \in {\mathbb{R}^{3}}$ is the sensor reading, ${}^{S}\mathrm{g}_{k}$ and ${}^{E}\mathrm{g}=[0,0,9.81]^{T}$ is the gravity vector on the sensor frame and earth frame, ${}^{S}\mathrm{a}_{k}$ is the free acceleration on the sensor frame, ${}_{E}^{S}q_k$ is the orientation (in a quaternion form) from the earth frame to the sensor frame, the operator $R({}_{E}^{S}q_k)$ converts the quaternion to a rotation matrix, and  $\mathrm{v}_{A,k}$ is the noise. The magnetometer model has 
  \begin{equation}
  	\begin{aligned}
  		\mathrm{y}_{M,k}&={}^{S}\mathrm{m}_{k}+{}^{S}\mathrm{d}_{k}+\mathrm{v}_{M,k}\\
  		{}^{S}\mathrm{m}_{k}&=R({}_{E}^{S}q_k){}^{E}\mathrm{m}
  		\label{magm}
  	\end{aligned}
  \end{equation}
  where $\mathrm{y}_{M,k}$ is the sensor reading, ${}^{S}\mathrm{m}_{k}$ and ${}^{E}\mathrm{m}$ is the earth magnetic vector on the sensor frame and earth frame (note that ${}^{E}\mathrm{m}$ is a constant and hence the subscript $k$ is ignored), ${}^{S}\mathrm{d}_{k}$ is the magnetic disturbance on the sensor frame, and $\mathrm{v}_{M,k}$ is the noise.
  
  As shown in \cite{b7}, the quaternion ${}_{E}^{S}q_k$ at the current step can be obtained by the quaternion at the precious step ${}_{E}^{S}q_{k-1}$ and the current angular velocity using
  \begin{equation}
  	\begin{aligned}
  		{}^{S}_{E}q_{k}&={}^{S}_{E}q_{k-1}+\frac{1}{2}
  		\left({}^{S}_{E}q_{k-1}\otimes \mathrm{w}_{k}^{q}\right)\Delta t
  		\label{gyrori}
  	\end{aligned}
  \end{equation}
  where $\Delta t$ is the sampling time, $\mathrm{w}_{k}^{q}=[0, \mathrm{w}_k^{T}]^{T}$ is the quaternion composed of true angular velocity $\mathrm{w}_k$. In a practical application, we use $\mathrm{y}_{G,k}$ to approximate $\mathrm{w}_{k}$, which gives the following equation
  \begin{equation}
  	{}^{S}_{E}q_{k}={}^{S}_{E}q_{k-1}+\frac{1}{2}
  	\left({}^{S}_{E}q_{k-1}\otimes \mathrm{y}_{G,k}^{q}\right)\Delta t + v_{G,k}
  	\label{gyrq}
  \end{equation}
  where $\mathrm{y}_{G,k}^{q}=[0, \mathrm{y}_{G,k}^{T}]^{T}$ and $v_{G,k}$ is the associated noise caused by the approximation of $\mathrm{w}_{k}$. Denoting $x_k={}^{S}_{E}q_{k}$, \eqref{gyrq} can be generalized as 
  \begin{equation}
  	x_k=f(x_{k-1},\mathrm{y}_{G,k})+ v_{G,k}
  	\label{pr}
  \end{equation}
  where $f(x_{k-1},\mathrm{y}_{G,k})=x_{k-1}+\left({}^{S}_{E}q_{k-1}\otimes \mathrm{y}_{G,k}^{q}\right)\Delta t$. Similarly, \eqref{accm} and \eqref{magm} can be reformulated as 
    \begin{equation}
  	y_k=h(x_k)+v_{AM,k}.
  	\label{mam}
  \end{equation}
  with 
  \begin{equation}\nonumber
  	y_k=\begin{bmatrix}
  		\mathrm{y}_{A,k}\\
  		\mathrm{y}_{M,k}
  	\end{bmatrix}, h(x_k)=\begin{bmatrix}
  	h_{A}(x_k)\\
  	h_{M}(x_k)
  	\end{bmatrix}, v_{AM,k}=\begin{bmatrix}
  	{v}_{A,k}\\
  	{v}_{M,k}
  	\end{bmatrix}
  \end{equation}
  where $h_{A}(x_k)=-R(x_k){}^{E}\mathrm{g}$, $h_{M}(x_k)=R(x_k){}^{E}\mathrm{m}$, ${v}_{A,k}={}^{S}\mathrm{a}_{k}+\mathrm{v}_{A,k}$, and ${v}_{M,k}={}^{S}\mathrm{d}_{k}+\mathrm{v}_{M,k}$. Equations \eqref{pr} and \eqref{mam} provide a general nonlinear state space model for IMUs. We then discuss the existing orientation estimation algorithms under the framework of \emph{maximum a posteriori} (MAP) and \emph{maximum likelihood estimation} (MLE). 
  
  Denote measurement set as $\mathbb{M}_{1:N}=\{\mathrm{y}_{A,k},  \mathrm{y}_{M,k}\}_{k=1}^{N}$, gyroscope reading set as $\mathbb{G}_{1:N}=\{\mathrm{y}_{G,k}\}_{k=1}^{N}$, and assume that $v_{G,k}$, $v_{AM,k}$ and the noise distribution for the initial guess $x_0$ are mutually independent. Under the MAP, the posterior distribution of $x_{1:N}$ given $\mathbb{M}_{1:N}$ [note that we regard $\mathbb{G}_{1:N}$ as known inputs of \eqref{pr}] has
  \begin{equation}
  \small
  \begin{aligned}
  	&p(x_{1:N}|\mathbb{M}_{1:N})=\frac{p(\mathbb{M}_{1:N}|x_{1:N})p(x_{1:N})}{p(\mathbb{M}_{1:N})}\\
	  	&=\frac{p(x_0)\prod_{k=1}^{N}p(y_k|x_k)\prod_{k=1}^{N}p(x_{k}|x_{k-1})}{p(\mathbb{M}_{1:N})}\\
  	&\propto p(x_0)\prod_{k=1}^{N}p_{v_{AM,k}}(y_k-h(x_k))\prod_{k=1}^{N}p_{v_{G,k}}(x_{k}-f(x_{k-1},\mathrm{y}_{G,k}))
  	\label{map}
  \end{aligned}
  \end{equation}
  where $p_{v_{AM,k}}(\cdot)$ and $p_{v_{G,k}}(\cdot)$ denotes the corresponding densities of $v_{AM,k}$ and $v_{G,k}$, respectively. By assuming that $v_{AM,k}\sim \mathcal{N}(0,R)$ (we will disclose \textbf{this assumption is inappropriate} in the following part), $v_{G,k}\sim \mathcal{N}(0,Q)$, $x_0\sim \mathcal{N}(\mu,\Pi)$ and ignoring the normalizing constant, the posterior distribution is given by
  \begin{equation}
  	\footnotesize	
  	\begin{aligned}
  	\exp&\left(-\frac{1}{2}\|\Pi^{-1/2}(x_0-\mu)\|_2^{2}\right)\prod_{k=1}^{N}\exp\left(-\frac{1}{2}\|R^{-1/2}(y_k-h(x_k))\|_2^{2}\right)\\
  	&\times\prod_{k=1}^{N}\exp\left(-\frac{1}{2}\|Q^{-1/2}(x_{k}-f(x_{k-1},\mathrm{y}_{G,k}))\|_2^{2}\right).
  	\end{aligned}
  	\label{mape}
  \end{equation}
  Maximizing \eqref{mape} is equivalent to minimizing its negative logarithm, which gives
  \begin{equation}
  \small
  \begin{aligned}
  	x_{1:N}&= \arg \min_{x_{1:N}} \frac{1}{2}\|\Pi^{-1/2}(x_0-\mu)\|_2^{2}\\
  	&+\frac{1}{2}\sum_{k=1}^{N}\|R^{-1/2}(y_k-h(x_k))\|_2^{2}\\
  	&+\frac{1}{2}\sum_{k=1}^{N}\|Q^{-1/2}(x_{k}-f(x_{k-1},\mathrm{y}_{G,k}))\|_2^{2}.
  \end{aligned}
  \label{mapobj}
  \end{equation}
  This optimization problem can be solved by extended Rauch–Tung–Striebel smoother~\cite{c11} or Gauss-Newton optimization~\cite{c12}.
  
  Under the MLE, by assuming that gyroscope readings $\mathrm{y}_{G,k}$ is independent with measurements $y_k$, the likelihood of observing the data $\{\mathbb{G}_{1:N},\mathbb{M}_{1:N}\}$ given $x_{1:N}$ has
  \begin{equation}
  \begin{aligned}
  	&x_{1:N}= \arg \max p(\{\mathbb{G}_{1:N},\mathbb{M}_{1:N}\}|x_{1:N})\\
  	&=p(\mathbb{G}_{1:N}|x_{1:N})p(\mathbb{M}_{1:N}|x_{1:N})\\
  	&=\prod_{k=1}^{N}p_{v_{G,k}}(x_{k}-f(x_{k-1},\mathrm{y}_{G,k}))\prod_{k=1}^{N}p_{v_{AM,k}}(y_k-h(x_k)).
  \end{aligned}
  \label{mle}
  \end{equation} 
  Taking negative logarithm on the right side of \eqref{mle} gives
  \begin{equation}
  	\small
  	\begin{aligned}
  	x_{1:N}&= \arg \min \frac{1}{2}\sum_{k=1}^{N}\|R^{-1/2}(y_k-h(x_k))\|_2^{2}\\
  	&+\frac{1}{2}\sum_{k=1}^{N}\|Q^{-1/2}(x_{k}-f(x_{k-1},\mathrm{y}_{G,k}))\|_2^{2}
  	\end{aligned}
  	\label{mleobj}
  \end{equation}
  which can be seen as a nonlinear regression problem with parameter vector $x_{1:N}$. Solutions for this problem include the Levenberg-Marquardt algorithm~\cite{c17}, Gauss-Newton optimization~\cite{c12}, etc.  
  
  In many applications, we are much more interested in the current state $x_k$ based on all previous measurements. Then, under the MAP, the filtering problem becomes
  \begin{equation}
  	\begin{aligned}
  		x_k &=\arg \min_{{x}_{k}}\frac{1}{2}\|R^{-1/2}(y_k-h(x_k))\|_{2}^{2}\\
  		&+\frac{1}{2}\|(P_{k}^{-})^{-1/2}({x}_{k}-f(x_{k-1},\mathrm{y}_{G,k}))\|_{2}^{2}
  		\label{ekf}
  	\end{aligned}
  \end{equation}
  where $P_{k}^{-}$ is the \emph{a priori} estimate of error covariance. Using \eqref{ekf} as an objective function, one can derives the extended Kalman filter~\cite{c14,c15}. Under the MLE, the filtering problem becomes 
  \begin{equation}
  	\begin{aligned}
  		x_k &=\arg \min_{{x}_{k}}\frac{1}{2}\|R^{-1/2}(y_k-h(x_k))\|_{2}^{2}\\
  		&+\frac{1}{2}\|Q^{-1/2}({x}_{k}-f(x_{k-1},\mathrm{y}_{G,k}))\|_{2}^{2}.
  		\label{gdm}
  	\end{aligned}
  \end{equation}
  One solution of \eqref{gdm} is the gradient descent method. Taking partial derivative on the right side of \eqref{gdm} with respect to $x_k$ and setting it to zero gives
  \begin{equation}\nonumber
  	-\left[\frac{\partial h}{\partial x}\Big|_{x_{k}}\right]R^{-1}(y_k-h(x_k))+Q^{-1}(x_k-f(x_{k-1},\mathrm{y}_{G,k}))=0
  \end{equation}
  Denoting the \emph{a priori} estimate of the state as $x_k^{-}=f(x_{k-1},\mathrm{y}_{G,k})$ and approximating $h(x_k)$ as $h(x_k^{-})$, it follows that
  \begin{equation}
  	x_k=x_k^{-}+Q\Big[\frac{\partial h}{\partial x}\Big|_{x_{k}^{-}}\Big]R^{-1}(y_k-h(x_{k}^{-})).
  	\label{gds}
  \end{equation}
  Assuming that $Q$ and $R$ are diagonal with the same diagonal entities with
  \begin{equation}
	Q=\operatorname{diag}([q,\ldots,q]), {R}=\operatorname{diag}([{r},\ldots,{r}]),
	\label{qr}
  \end{equation}
  it follows that
  \begin{equation}
  	x_k=x_k^{-}+\lambda\Big[\frac{\partial h}{\partial x}\Big|_{x_{k}^{-}}\Big](y_k-h(x_{k}^{-}))
  	\label{gds1}
  \end{equation}
  with $\lambda=\frac{q}{r}$ determining how confident we are in believing the measurements compared with the prediction. It is worth mentioning some authors utilize $h(x_{k-1})$ to approximate $h(x_k)$~\cite{b7}, which gives
   \begin{equation}
  	x_k=x_k^{-}+\lambda\Big[\frac{\partial h}{\partial x}\Big|_{x_{k-1}}\Big](y_k-h(x_{k-1})).
  \end{equation}  
  This small difference usually does not have a big impact on the algorithm performance of IMUs since both $x_{k-1}$ and $x_{k}^{-}$ provide a good initial guess of $x_k$. 
  
  An alternative is to use the inverse of the nonlinear measurement model rather than the original model in problem formulation, i.e., constructing $x_k = h^{-1}(y_k) + \breve{v}_{AM,k}$ and assuming $\breve{v}_{AM,k} \sim \mathcal{N}(0,\breve{R})$. By this conversion, \eqref{gdm} can be rewritten as 
  \begin{equation}
  	\begin{aligned}
  		x_k &=\arg \min_{{x}_{k}}\frac{1}{2}\|\breve{R}^{-1/2}(x_k-h^{-1}(y_k))\|_{2}^{2}\\
  		&+\frac{1}{2}\|Q^{-1/2}({x}_{k}-f(x_{k-1},\mathrm{y}_{G,k})\|_{2}^{2}.
  	\end{aligned}
  	\label{cm}
  \end{equation}
  Taking partial derive on the right side of \eqref{cm} and setting it to zero gives
  \begin{equation}
  	x_k=[Q^{-1}+\breve{R}^{-1}]^{-1}Q^{-1}x_k^{-}+[Q^{-1}+\breve{R}^{-1}]^{-1}\breve{R}^{-1}h^{-1}(y_k).
  \end{equation}
   If $Q$ and $\breve{R}$ are diagonal matrices with the same entities on the main diagonal, i.e., $Q=\operatorname{diag}([q,\ldots,q])$ and $\breve{R}=\operatorname{diag}([\breve{r},\ldots,\breve{r}])$, the above equation can be reformulated as
   \begin{equation}
   	x_k= \gamma x_k^{-} + (1-\gamma)x_{AM,k}
   	\label{cmf}
   \end{equation}
   where $\gamma=\frac{\breve{r}}{q+\breve{r}}$ and $x_{AM,k}$ is the state determined by measurement $y_k$ with $x_{AM,k}=h^{-1}(y_k)$. Equation \eqref{cmf} gives a general form of complementary filter for orientation estimation~\cite{c16}. A remaining question is the obtainment of $x_{AM,k}$. One method is to solve the following LS-based objective function:
   \begin{equation}
   	x_{AM,k} =\arg\min \frac{1}{2}\|y_k-h(x_{AM,k})\|_2^{2}.
   	\label{lsobj}
   \end{equation} 
   Applying one-step gradient descent update with the initial guess of  $x_{AM,k}$ as $x_k^{-}$ gives
   \begin{equation}
   	x_{AM,k} = x_{k}^{-} + \mu\Big[\frac{\partial h}{\partial x}\Big|_{x_{k}^{-}}\Big]\left(y_k-h(x_{k}^{-})\right).
   	\label{xam}
   \end{equation}
   where $\mu$ is the learning rate. Substituting \eqref{xam} into \eqref{cmf}, one obtains
   \begin{equation}
   	\begin{aligned}
   	   	x_k&= \gamma x_k^{-} + (1-\gamma)\left(x_{k}^{-} + \mu\Big[\frac{\partial h}{\partial x}\Big|_{x_{k}^{-}}\Big]\left(y_k-h(x_{k}^{-})\right)\right)\\
   	   	&=x_k^{-} + \mu(1-\gamma)\Big[\frac{\partial h}{\partial x}\Big|_{x_{k}^{-}}\Big]\left(y_k-h(x_{k}^{-})\right).
   	\end{aligned}
   	\label{cfs}
   \end{equation}
   One can observe that \eqref{cfs} is identical with \eqref{gds1} by assigning $\mu(1-\gamma)=\lambda$. This indicates that the optimization problems \eqref{gdm} and \eqref{cm} are interchangeable under certain situations and the key of \eqref{cm} is solving \eqref{lsobj}. 
  
\begin{figure}[htbp]
  	\centering
  	\subfigure[without disturbance]{
  		\begin{minipage}[t]{0.5\linewidth}
  			\centering
  			\includegraphics[width=1\columnwidth]{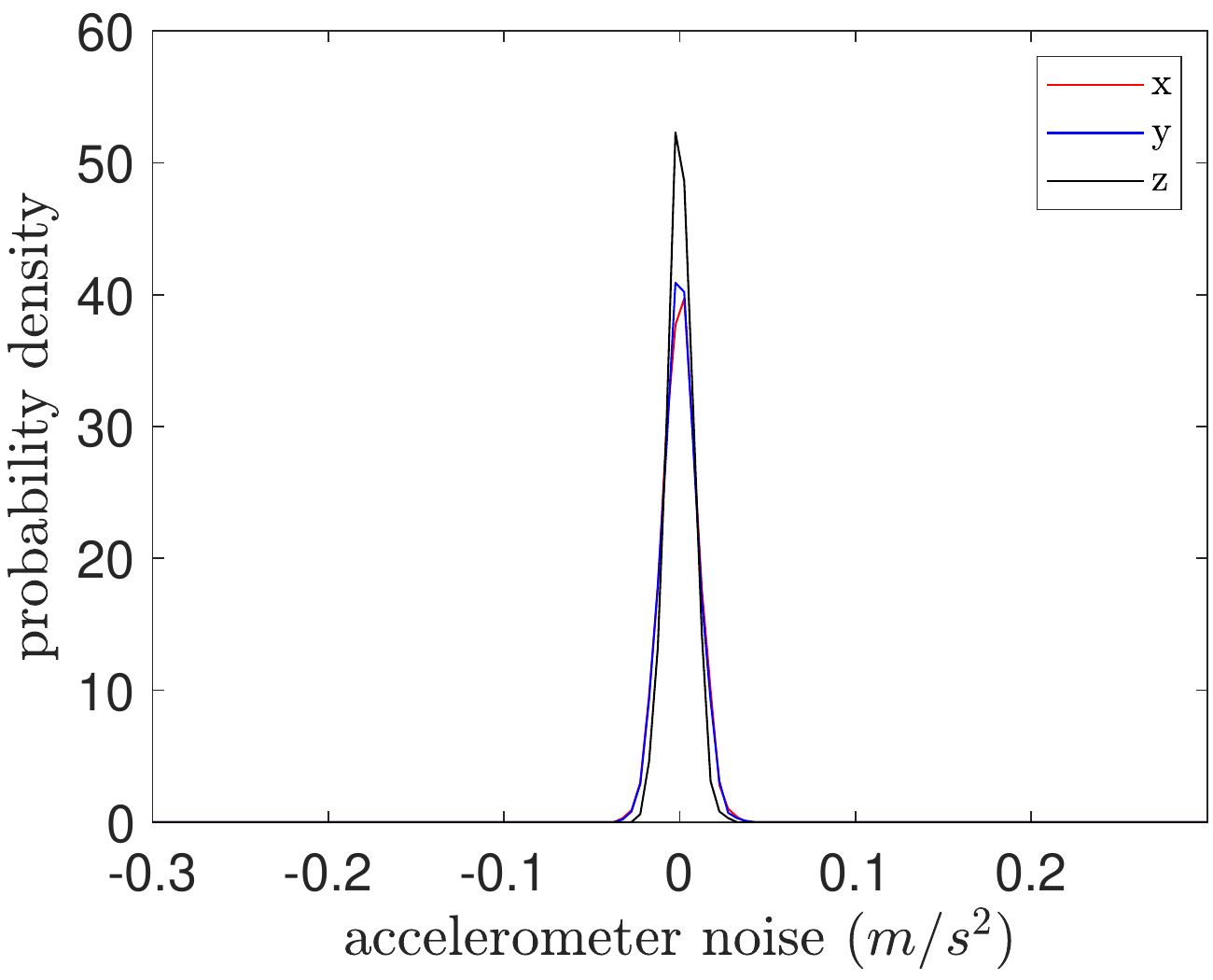}
  			\label{acc_nd}	
  		\end{minipage}%
  	}%
  	\subfigure[with disturbance]{
  		\begin{minipage}[t]{0.51\linewidth}
  			\centering
  			\includegraphics[width=1\columnwidth]{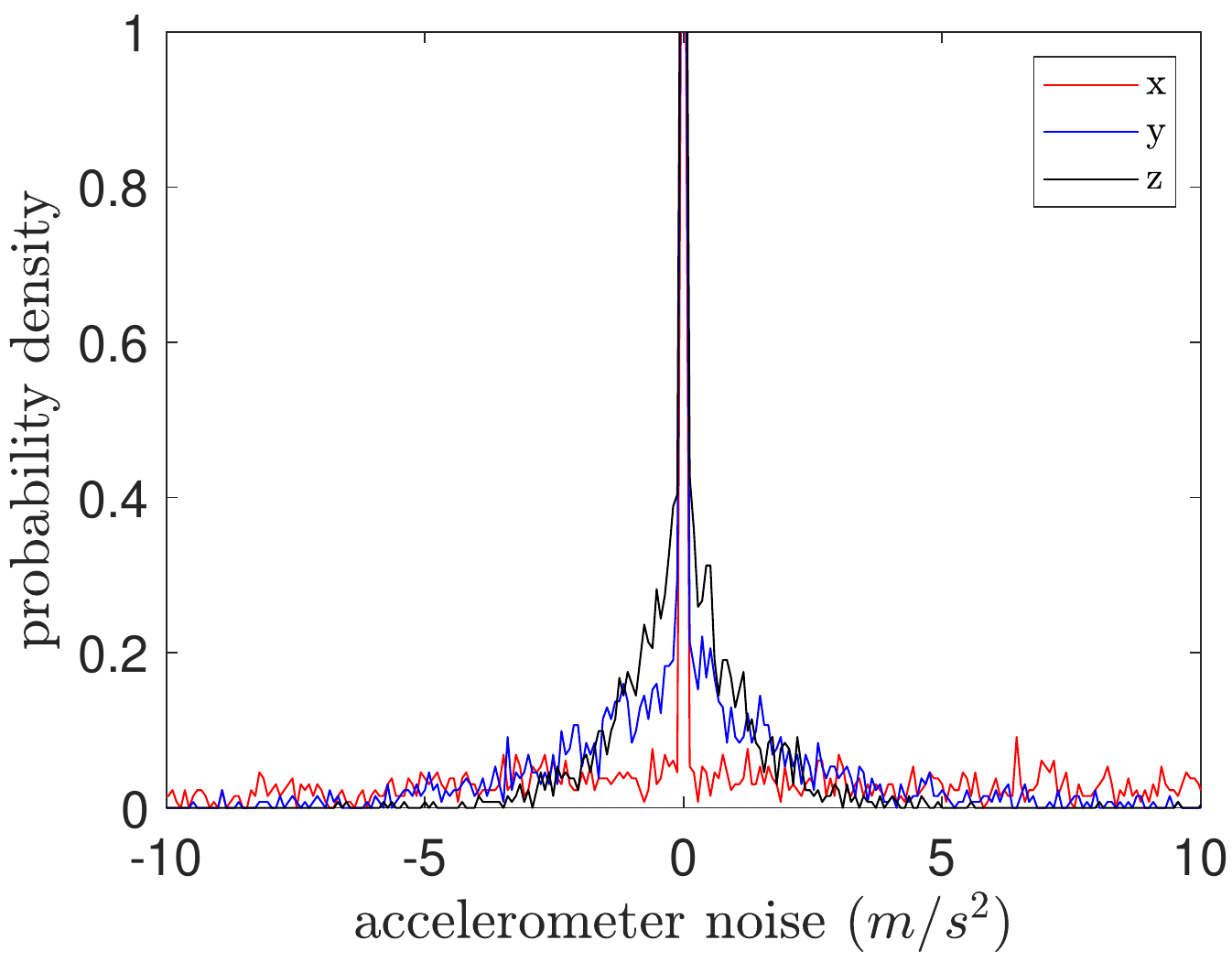}
  			\label{acc_wd}	
  		\end{minipage}%
  	}%
  	\\
  	\subfigure[without disturbance]{
  		\begin{minipage}[t]{0.50\linewidth}
  		\centering
  		\includegraphics[width=1\columnwidth]{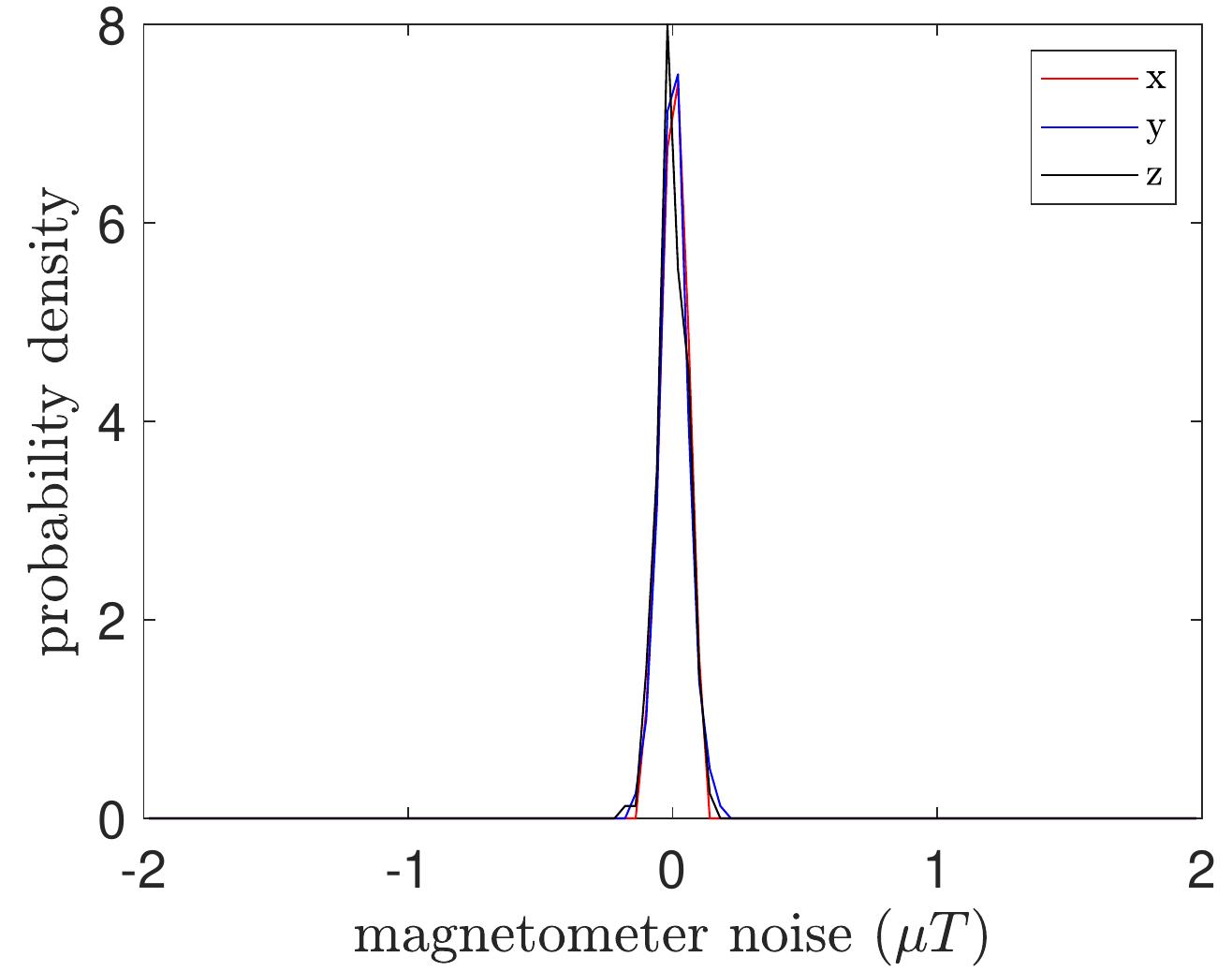}
  		\label{mag_nd}	
  	\end{minipage}%
  	}%
  	\subfigure[with disturbance]{
  	\begin{minipage}[t]{0.50\linewidth}
  		\centering
  		\includegraphics[width=1\columnwidth]{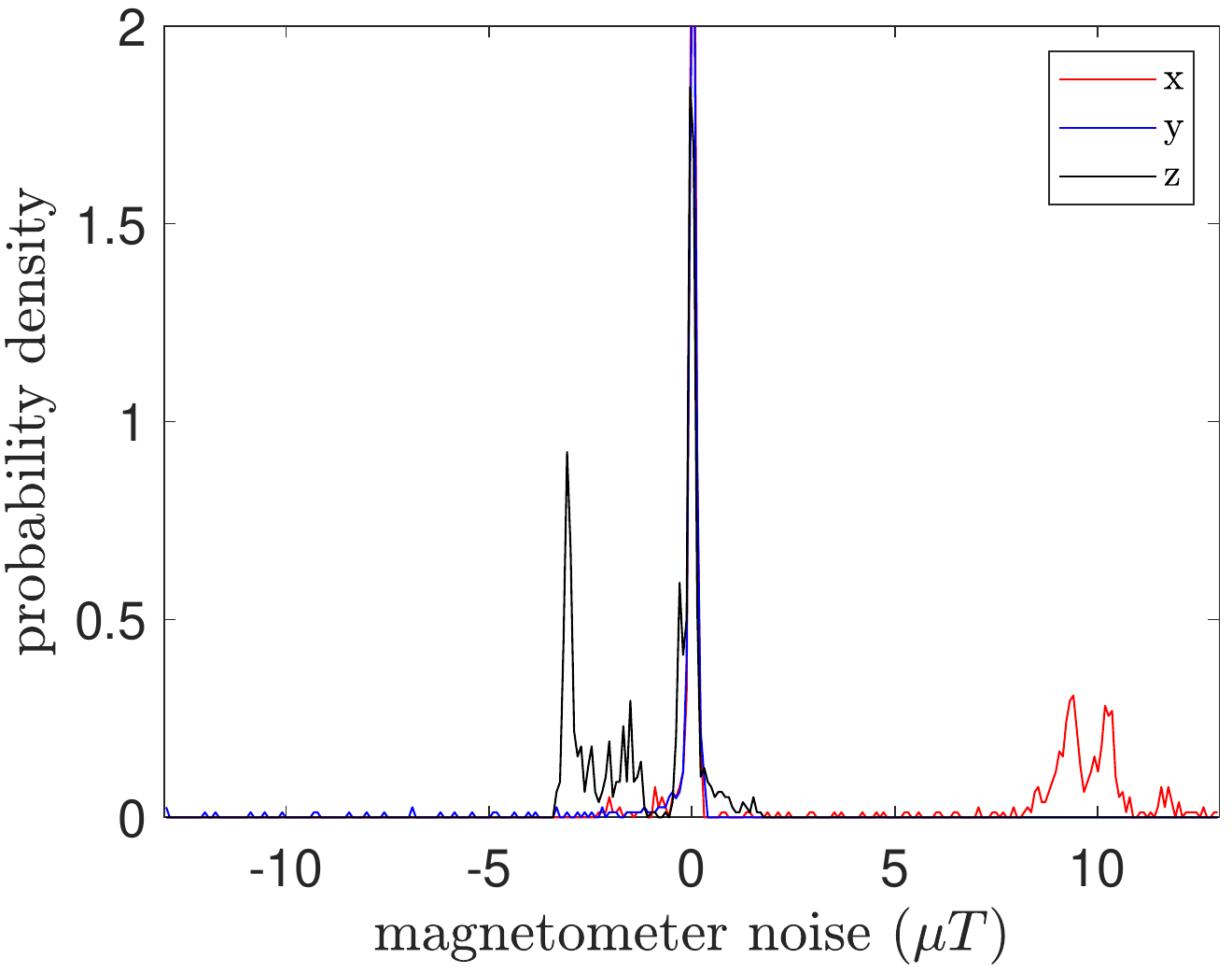}
  		\label{mag_wd}	
  	\end{minipage}%
  	}%
  	\caption{\textcolor{black}{The pdfs of $v_{A,k}$ and $v_{M,k}$ without and with disturbances. (a) and (c) show the pdfs of $v_{A,k}$ and $v_{M,k}$ when IMU is static and free of disturbances. (b) and (d) show the pdfs of $v_{A,k}$ and $v_{M,k}$ when IMU is with disturbances. The IMU used in this experiment is Xsens MTI-670 and the disturbed data is generated by the translation experiment as shown in Fig. \ref{linear}.}}		
  	\label{accn}
  \end{figure}

  From \eqref{ekf}, \eqref{gdm}, and \eqref{cm}, one can see that conventional orientation estimation algorithms rely heavily on the Gaussian assumptions~\cite{b6,b7,b14,c16}. This assumption may be valid for gyroscope noise $v_{G,k}$, but not for accelerometer noise $v_{A,k}$ and magnetometer noise $v_{M,k}$ due to the existence of unknown acceleration ${}^{S}\mathrm{a}_k$ and magnetic disturbance ${}^{S}\mathrm{d}_k$~\cite{c15}. A better density representation for $v_{A,k}$ and $v_{M,k}$ should be the heavy-tailed distribution. To illustrate this, we investigate the probability density functions (pdfs) of $v_{A,k}$ and $v_{M,k}$ in two situations: without and with disturbances. We first keep the IMU being static and free of disturbances and the corresponding pdfs of $v_{A,k}$ and $v_{M,k}$ are shown in Figs. \ref{acc_nd} and \ref{mag_nd}. Then, we manually generate external acceleration and magnetic disturbance, and the corresponding results are shown in Figs. \ref{acc_wd} and \ref{mag_wd}. The disturbance generation method is described in the caption of Fig. \ref{linear} (hence is ignored here). Not surprisingly, The densities of $v_{A,k}$ and $v_{M,k}$ are Gaussian-like when without disturbances and are heavy-tailed when involving disturbances. Unfortunately, in many practical applications of IMUs, external acceleration and magnetic disturbance are unavoidable, which degenerates the conventional algorithms significantly.
  
  \subsection{\textcolor{black}{Properties of the MKCL as a Cost Function}} 
  In this section, we discuss the properties of the MKCL as an objective function from the perspective of noise distribution, likelihood function, and influence function.
   
  The MLE viewpoint in \eqref{mle} can be easily extended to the general case of noise distributions, i.e., 
  \begin{equation}\nonumber
  	\small
  	\begin{aligned}
  		p(v_{AM}) &\propto \exp(-J_r(R^{-1/2}v_{AM}))\\
  		p(v_{G}) &\propto \exp(-J_q(Q^{-1/2}v_{G}))
  	\end{aligned}
  \end{equation}
  where $J_{r}(\cdot)$ and $J_{q}(\cdot)$ are nonlinear functions with respect to its argument. Then, based on \eqref{mle}, we can replace the mean squared errors $\|R^{-1/2}v_{AM}\|_2^{2}$ and $\|Q^{-1/2}v_{G}\|_2^{2}$ by nonlinear functions $J_r(R^{-1/2}v_{AM}$ and $J_q(Q^{-1/2}v_{G})$ and construct the following problem:
  \begin{equation}
  	\small
  	\begin{aligned}
  		x_{1:N}&= \arg \min \sum_{k=1}^{N}J_r\Big(R^{-1/2}\big(y_k-h(x_k)\big)\Big)\\
  		&+\sum_{k=1}^{N}J_q\Big(Q^{-1/2}\big(x_{k}-f(x_{k-1},\mathrm{y}_{G,k})\big)\Big).
  	\end{aligned}
  	\label{mlemkcl}
  \end{equation}
The corresponding filtering problem becomes:
  \begin{equation}
  	\small
  	\begin{aligned}
  		x_k&= \arg \min J_r\Big(R^{-1/2}\big(y_k-h(x_k)\big)\Big)\\
  		&+J_q\Big(Q^{-1/2}\big(x_{k}-f(x_{k-1},\mathrm{y}_{G,k})\big)\Big).
  	\end{aligned}
  	\label{mlefmkcl}
  \end{equation}
    
  A good candidate for the nonlinear $J_{r}(\cdot)$ and $J_{q}(\cdot)$ is the MKCL. To investigate the properties of the MKCL as a cost function, we consider the following nonlinear regression problem:
  \begin{equation}
  	y_k = g(x,u_k) + e_k 
  	\label{regression}
  \end{equation}
  where $y_k \in \mathbb{R}^{l}$ is the output, $u_k$ is the input, $x$ is the parameter vector to be estimated, $g(\cdot)$ is the nonlinear function, $e_k$ is the noise, and $k=1,2,\ldots,~N$ is the sample index. We assume that the nominal distribution of $e_k$ has $e_k \sim \mathcal{N}(0,R)$. Then, we compare the following two objective functions:  
   \begin{subequations}
   	 	\label{cost}
 	\begin{align}
 		\arg \min_{x} J_{LS}(\tilde{e}_k) = \frac{1}{2N}\sum_{k=1}^{N} \sum_{i=1}^{l}\tilde{e}_{i,k}^{2}
 		\label{mseobj}
 	\end{align}
 	\begin{align}
 		\arg \min_x J_{CL}(\tilde{e}_k) = \frac{1}{N}\sum_{k=1}^{N}\sum_{i=1}^{l}\sigma_i^2 \Big(1- G_{\sigma_i}(\tilde{e}_{i,k})\Big)
 		\label{clobj}
 	\end{align}
 \end{subequations} 
  where $\tilde{e}_k \triangleq R^{-1/2}(y_k-g(x,u_k))$ and $\tilde{e}_k=[\tilde{e}_{1,k},\tilde{e}_{2,k},\ldots,\tilde{e}_{l,k}]^{T}$. Then, we have the following theorem.
  \color{black}
   \begin{theorem}
  	$J_{LS}$ in \eqref{mseobj} is an optimal loss function when $\tilde{e}_k \sim \mathcal{N}(0, I)$ based on MLE. On the contrary, $J_{CL}$ in \eqref{clobj} is an optimal cost function if $\tilde{e}_{i,k}$ follows 
  	\begin{equation}
  		p\big(\tilde{e}_{i,k}\big)=c_i \exp\Big(-\sigma_i^2(1-\exp\big(-\frac{\tilde{e}_{i,k}^2}{2\sigma_i^2})\big)\Big)
  		\label{pdfe}
  	\end{equation}
  	where $c_i$ is a normalization coefficient and $i=1,2,\ldots,l$.
  	\label{theorem2}
  \end{theorem}
  
\begin{figure}[htbp]
  	\centerline{\includegraphics[width=5.5cm]{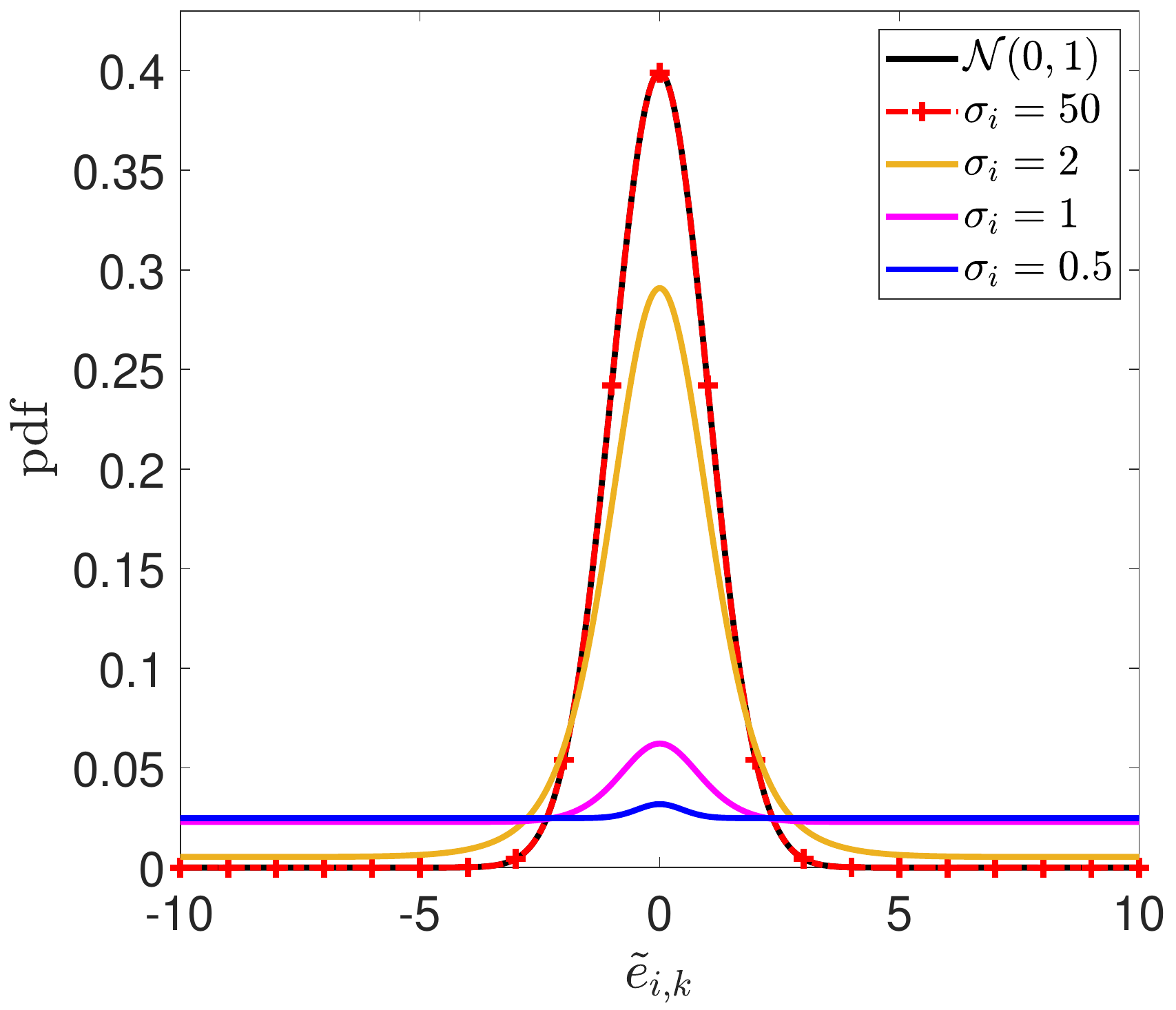}}
  	\caption{The pdfs of  $\mathcal{N}(0,1)$ and $p(\tilde{e}_{i,k})$ in \eqref{pdfe} with different kernel bandwidths. The error $\tilde{e}_{i,k}$ is assumed to be bounded within $[-20,20]$ for $p(\tilde{e}_{i,k})$ and the coefficient $c_i$ is obtained by $c_i=1/\int_{-20}^{20} \exp\Big(-\sigma_i^2\big(1-\exp(-\frac{\tilde{e}_{i,k}^2}{2\sigma_i^2})\big)\Big) \mathrm{d} \tilde{e}_{i,k}$.}
  	\label{pdf_ek}
  \end{figure}

  The proof of this theorem is shown in Appendix \ref{proof2}. A comparison of Gaussian distribution $\mathcal{N}(0,1)$ and $p(\tilde{e}_{i,k})$ in \eqref{pdfe} with different kernel bandwidth is shown in Fig. \ref{pdf_ek}. As depicted, when $\sigma_i$ is large, $p(\tilde{e}_{i,k})$ approaches a Gaussian distribution, which aligns with Theorem \ref{theorem1}. When $\sigma_i$ is relatively small, it represents a type of heavy-tailed distribution. This indicates that the MKCL is a good candidate when the noise is heavy-tailed.
  
\begin{figure}[htbp]
   	\centerline{\includegraphics[width=6.5cm]{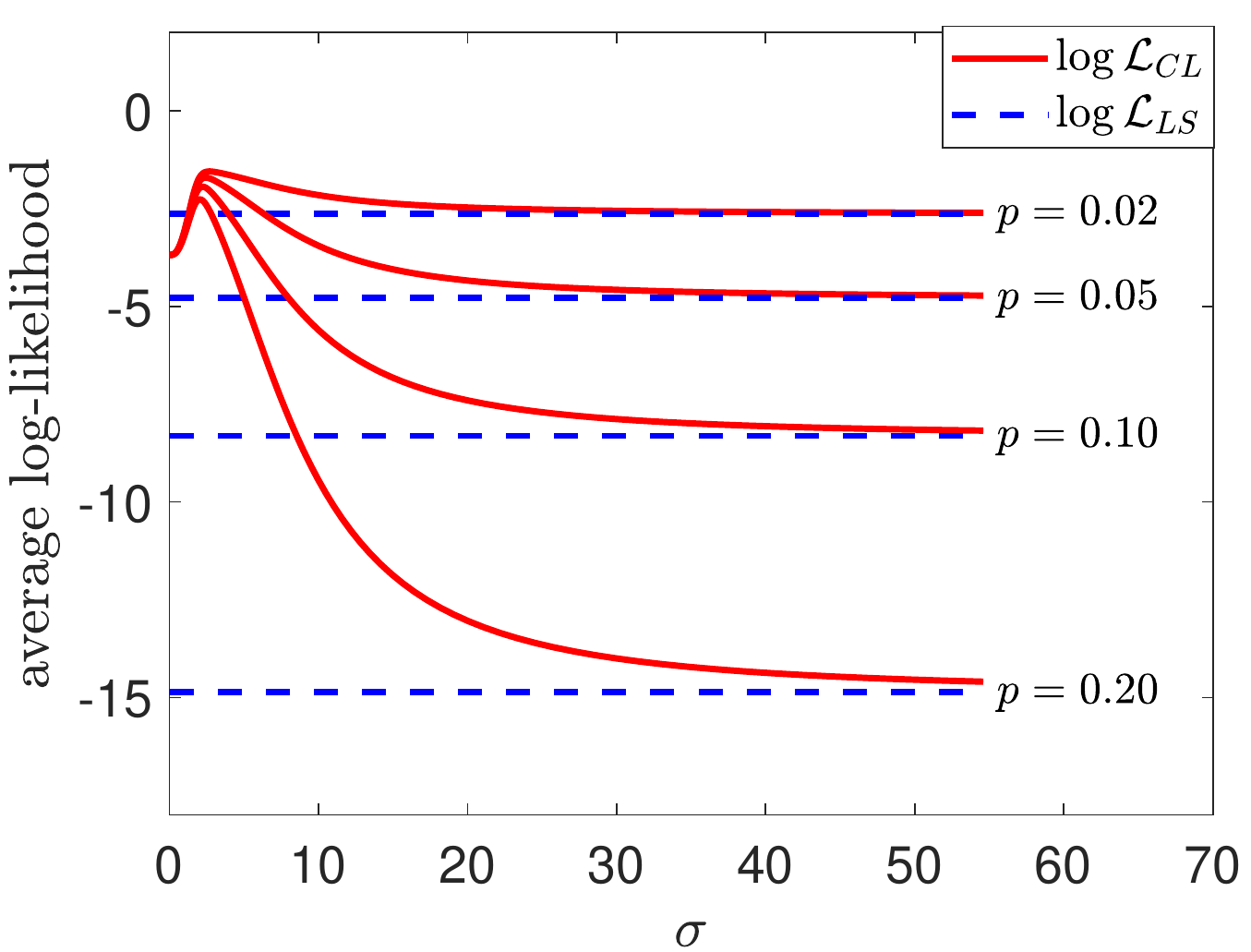}}
   	\caption{A comparison of $\log \mathcal{L}_{CL}$ and $\log \mathcal{L}_{LS}$ under differ $p$.}
   	\label{likelihood}
   \end{figure}

   \color{black}
   To further explore the properties of the MKCL as a loss function, we use the log-likelihood function as a metric to investigate how well different objective functions explains the data. We consider the following mixture noise distribution with
   \begin{equation}
    \tilde{e}_k \sim (1-p)\mathcal{N}(0,1) + p\mathcal{U}(-20,20), 0 \le p<0.5
   	\label{residual}
   \end{equation}
   where $\mathcal{N}(0,1)$ is the nominal Gaussian distribution, $\mathcal{U}(-20,20)$ is a uniform distribution with boundary $[-20,20]$, and $p$ is a probability that determines $\tilde{e}_k$ generated by which distribution. The log-likelihood for the MKCL can be obtained by $\log\mathcal{L}_{CL}=\frac{1}{N}\sum_{k=1}^{N}p(\tilde{e}_k)$ where $p(\tilde{e}_k)$ is shown in \eqref{pdfe} and the value for the LS has $\log\mathcal{L}_{LS}=\frac{1}{N}\sum_{k=1}^{N}\frac{1}{\sqrt{2\pi}}\exp\left(-\frac{\tilde{e}_{k}^2}{2}\right)$. We compare $\log \mathcal{L}_{CL}$ and $\log \mathcal{L}_{LS}$ under different $p$. The results are shown in Fig.\ref{likelihood}. We observe that $\mathcal{L}_{CL} > \mathcal{L}_{LS}$ always holds when $\sigma$ is bigger than a certain threshold $\sigma^{*}$ and $p>0$, which indicates that the MKCL is much more preferable when the noise is heavy-tailed. We also observe that the potential benefits of the MKCL over the LS [i.e., maximum values of $(\log \mathcal{L}_{CL}-\log\mathcal{L}_{LS})$] increases with the growth of $p$, which implies that a higher profit can be obtained by replacing the LS with the MKCL when the noise has a heavier tail. This fact also reveals that the LS criterion is sensitive to heavy-tailed noises but the MKCL is not, especially when the kernel bandwidth is properly selected.
  
\begin{figure}[htbp]
  	\centering
  	\subfigure[Objective function]{
  		\begin{minipage}[t]{0.49\linewidth}
  			\centering
  			\includegraphics[width=1\columnwidth]{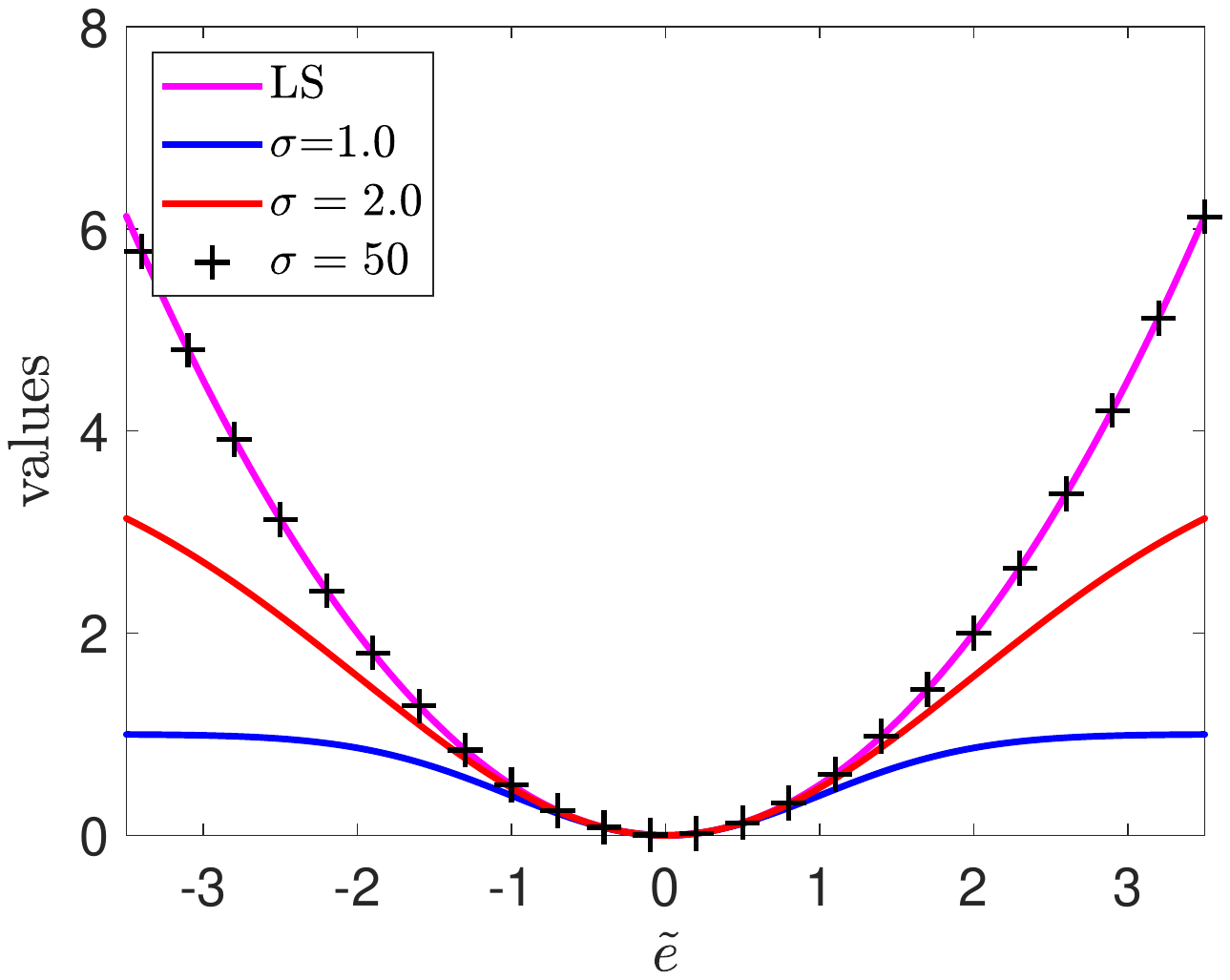}
  			\label{obj}	
  		\end{minipage}%
  	}%
  	\subfigure[Influence function]{
  		\begin{minipage}[t]{0.50\linewidth}
  			\centering
  			\includegraphics[width=1\columnwidth]{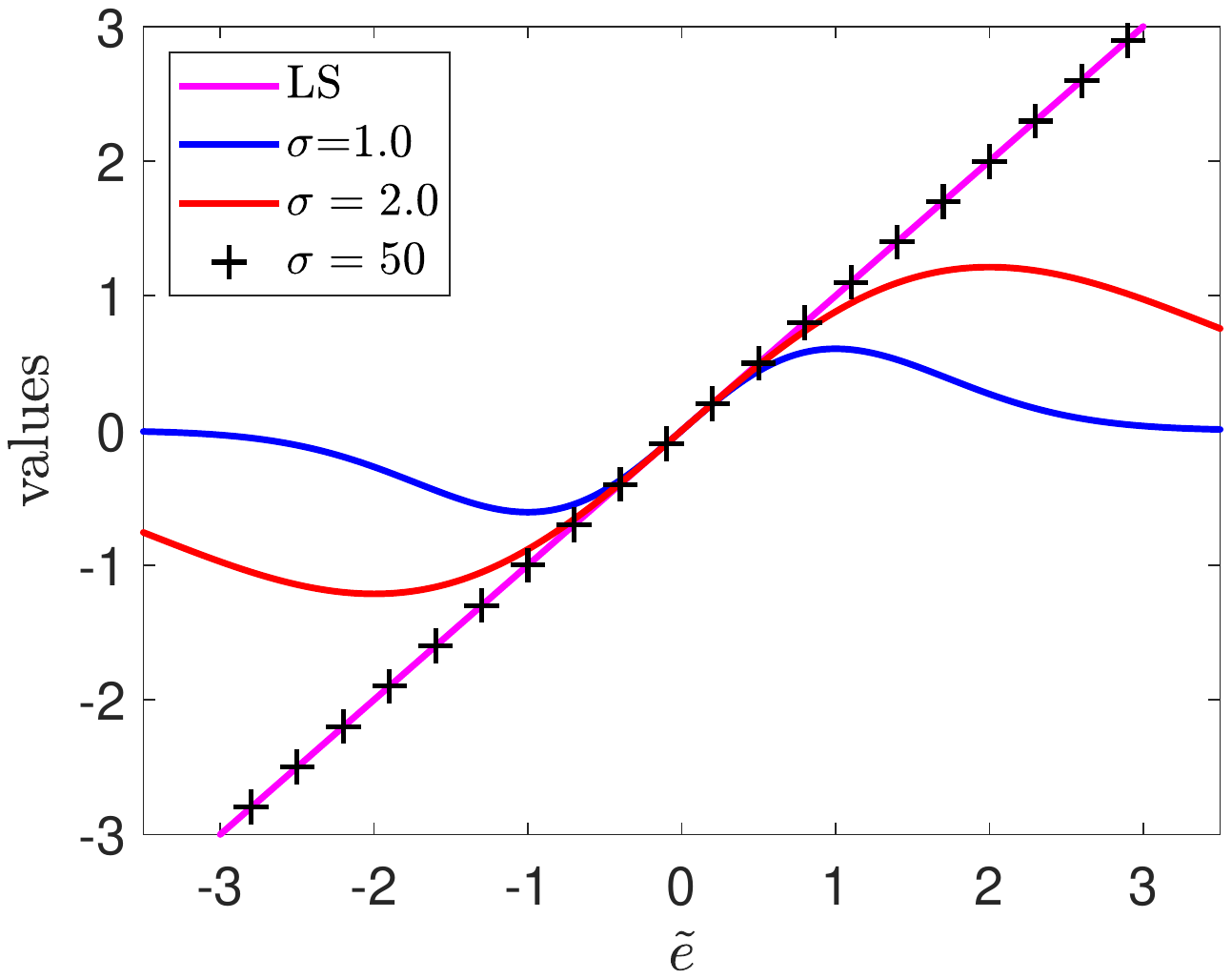}
  			\label{inf}	
  		\end{minipage}%
  	}%
  	\caption{Objective functions and influence functions for $J_{LS}$ and $J_{CL}$ with different kernel bandwidths.}		
  	\label{influence}
  \end{figure}

  The influence function measures the derivative of the loss with respect to the residual and quantifies the effect of the size of a residual on the loss~\cite{c18}. In the case of $l=1$ and $N=1$ in \eqref{cost}, one has $J_{LS}=\frac{1}{2}\tilde{e}^2$ and $J_{CL}=\sigma^2(1-G_{\sigma}(\tilde{e}))$. Then, we visualize the objective function and influence of $J_{LS}$ and $J_{CL}$ with different kernel bandwidths in Fig. \ref{influence}. One can see that the LS gives each residual constant influence as shown in Fig. \ref{inf}, which indicates that they are sensitive to outliers. On the contrary, the influence function of the MKCL is close to that of the LS when the residual is small, but goes towards zero with the growth of the residual (when using small kernel bandwidth). This redescending property makes it an attractive option when the underlying noise is heavy-tailed.
  
  \subsection{\textcolor{black}{Problem Formulation}}
  \label{formulation}
  As explained in Section \ref{sensormodel}, the Gaussian assumption is valid for gyroscopes but not for accelerometers and magnetometers. Taking this prior knowledge into consideration,  one can construct the following problem:
    \begin{equation}
  	\small
  	\begin{aligned}
  		x_k&= \arg \min_{x_k} J_{CL}\left(R^{-1/2}\big(y_k-h(x_k)\big)\right)\\
  		&+\|Q^{-1/2}\big(x_{k}-f(x_{k-1},\mathrm{y}_{G,k})\big)\|_{2}^{2}.
  	\end{aligned}
  	\label{mkcl}
  \end{equation}
  Denoting $\tilde{e}_k=R^{-1/2}\big(y_k-h(x_k)\big) \in \mathbb{R}^{l}$, it follows that $J_{CL}(\tilde{e}_k)=\sum_{i=1}^{l}\sigma_i^{2}\big(1-G_{\sigma_i}(\tilde{e}_{i,k})\big)$. By taking partial derivative on the right side of above equation with respect to $x_k$, one has
 \begin{equation}
 	\small
	-\Big[\frac{\partial h}{\partial x}\Big|_{x_{k}}\Big]\tilde{W}R^{-1}(y_k-h(x_k))+Q^{-1}(x_k-f(x_{k-1},\mathrm{y}_{G,k}))=0
 \end{equation}
 where $\tilde{W}=\operatorname{diag}([G_{\sigma_1}(\tilde{e}_{1,k}),G_{\sigma_2}(\tilde{e}_{2,k}),\ldots,G_{\sigma_l}(\tilde{e}_{l,k})])$. By analogy with \eqref{gds}, one obtains
   \begin{equation}
 	x_k=x_k^{-}+Q\Big[\frac{\partial h}{\partial x}\Big|_{x_{k}^{-}}\Big]\tilde{W}R^{-1}(y_k-h(x_{k}^{-})).
 	\label{gds2}
 \end{equation}
  
  A much more straightforward method is to follow the procedure shown in \eqref{cm} to \eqref{cfs} and construct the following problem
  \begin{equation}
  	x_{AM,k} =\arg\min J_{CL}\left(y_k-h(x_{AM,k})\right).
  	\label{lsobj1}
  \end{equation} 
  Denoting $e_k=y_k-h(x_{k}^{-}) \in  \mathbb{R}^{l}$ and by analogy with \eqref{xam}, it follows that
  \begin{equation}
  	x_{AM,k} = x_{k}^{-} + \mu\Big[\frac{\partial h}{\partial x}\Big|_{x_{k}^{-}}\Big]W\left(y_k-h(x_{k}^{-})\right)
  	\label{xam1}
  \end{equation}
   where $e_k=[e_{1,k},e_{2,k},\ldots,e_{l,k}]^{T}$ and $W=\operatorname{diag}([G_{\sigma_1}({e}_{1,k}),G_{\sigma_2}({e}_{2,k}),\ldots,G_{\sigma_l}({e}_{l,k})])$. Substituting \eqref{xam1} into \eqref{cmf} gives
	\begin{equation}
    x_k=x_k^{-} + \mu(1-\gamma)\Big[\frac{\partial h}{\partial x}\Big|_{x_{k}^{-}}\Big]W\left(y_k-h(x_{k}^{-})\right).
    \label{cfs2}
	\end{equation}
  
  One can observe that the only difference between \eqref{cfs2} and \eqref{cfs} is the attenuating matrix $W$. The utility of this matrix can be understood intuitively: when no external disturbances are involved in the measurements, one can expect that the discrepancy $y_k-h(x_{k}^{-})$ is small and hence $W \approx I$. In this case, equation \eqref{cfs2} is similar to \eqref{cfs}. On the contrary, with the presence of disturbances, the discrepancy $y_k-h(x_{k}^{-})$ is expected to be big, and hence $W$ is close to a zero matrix. By this technique, the MKCL-based methods are much more robust than their LS-based counterparts where the ``robustness level" is controlled by the kernel bandwidth.

  \begin{remark}
	It is worth mentioning that the purpose of Sections II-B and II-D is to provide a general framework for orientation estimation algorithms. In specific algorithms, different orientation parametrizing strategies, e.g., quaternions, rotation matrices, rotation vectors, and Euler angles, are employed by different authors in different applications. Moreover, some other variables, e.g., the gyroscope bias, the external disturbance, and the magnetic disturbance, may be augmented into the state to enhance the algorithm's performance. The purpose of this work is not to obtain the best orientation estimation algorithm but to demonstrate that an improvement can be further obtained by replacing the LS cost with the much more generic MKCL. The newly derived algorithms would inherit the advantages of the original method but with enhanced robustness to disturbances.
  \end{remark}
  \color{black}
 \section{Main Results}
  \label{algd}
  
  In this section, we first demonstrate that two popular algorithms, i.e., the GD and DOE, are derived under the LS-based objective functions. Then we use MKCL to substitute LS and derive two novel  algorithms, i.e., the CGD and CDOE. Before proceeding, we denote the normalized accelerometer readings as ${{y}}_{A,k}=\frac{{\mathrm{y}}_{A,k}}{\|{\mathrm{y}}_{A,k}\|_2}=[y_{ax},y_{ay},y_{az}]^{T}$ and the normalized magnetometer readings as  ${{y}}_{M,k}=\frac{{\mathrm{y}}_{M,k}}{\|{\mathrm{y}}_{M,k}\|_2}=[y_{mx},y_{my},y_{mz}]^{T}$.
\subsection{Two Traditional Solutions for IMUs}   
	\label{gddoe}
   In this section, we provide a brief overview of the derivation of the GD~\cite{b7} and DOE~\cite{b14}, emphasizing that they are derived using LS-based objective functions (note that this fact is not pointed out by their original authors). 
 	\subsubsection{The GD}
	The GD algorithm in ~\cite{b7} is composed of two steps: the \emph{a priori} estimate of orientation by gyroscope readings and the correction by accelerometer and magnetometer readings. Based on \eqref{gyrq}, the \emph{a priori} estimate of quaternion ${}^{S}_{E}q_{k}^{-}$ has
\begin{equation}
	{}^{S}_{E}q_{k}^{-}={}^{S}_{E}q_{k-1}+\frac{1}{2}
	\left({}^{S}_{E}q_{k-1}\otimes \mathrm{y}_{G,k}^{q}\right)\Delta t
	\label{gyrq1}
\end{equation}
where ${}^{S}_{E}q_{k-1}$ is the quaternion at previous time step, and $\mathrm{y}_{G,k}^{q}=[0, \mathrm{y}_{G,k}^{T}]^{T}$. For accelerometers, the constant gravity vector quaternion is set to be  ${}^{E}{g}=[0,0,0,1]^{T}$ and the measured gravity quaternion is ${}^{S}{g}=[0,{{y}}_{A,k}^{T}]^{T}$. The following LS-based objective function is utilized:
\begin{equation}
	\min_{{}^{S}_{E}q_{k} \in \mathbb{R}^{4}} f_g({}^{E}g, {}^{S}{g}, {}^{S}_{E}q_{k})=\frac{1}{2}\|E_{g,k}\|^{2}
	\label{acc}
\end{equation}
where $ E_{g,k}={}^{S}_{E}q_{k}^{*}\otimes {}^{E}g \otimes{}^{S}_{E}q_{k} - {}^{S}{g}$ is the error quaternion, ${}^{S}_{E}q_{k}=[q_1,q_2,q_3,q_4]^{T}$ is the quaternion to be determined, and ${}^{S}_{E}q_{k}^{*}$ is the conjugate quaternion of ${}^{S}_{E}q_{k}$. For magnetometers, the constant magnetic vector quaternion is set to be ${}^{E}m=[0, m_x,0,m_z]^{T}$ (due to magnetic dip) with $\sqrt{m_x^2+m_z^2}=1$ and the measured magnetic vector quaternion is ${}^{S}m=[0,{{y}}_{M,k}^{T}]^{T}$. The following LS-based cost function is employed:
\begin{equation}
	\min_{{}^{S}_{E}q_{k} \in \mathbb{R}^{4}} f_m({}^{E}m, {}^{S}m, {}^{S}_{E}q_{k})=\frac{1}{2}\|E_{m,k}\|^{2}
	\label{mag}
\end{equation}
where $E_{m,k}={}^{S}_{E}q_{k}^{*}\otimes {}^{E}m \otimes{}^{S}_{E}q_{k} - {}^{S}m$. Combining \eqref{acc} and \eqref{mag}, one has
	\begin{equation}
	\min_{{}^{S}_{E}q_{k} \in \mathbb{R}^{4}} f_{g,m}({}^{E}g_{}, {}^{E}m_{}, {}^{S}g_{}, {}^{S}m_{}, {}^{S}_{E}q_{k})=\frac{1}{2}\norm{E_{k}}^{2}
	\label{accmag}
	\end{equation}
	where $E_{k}=\begin{bmatrix}
	E_{g,k}\\
	E_{m,k}
	\end{bmatrix}\in \mathbb{R}^{8}$ is the aggregated error quaternion. By  applying one step gradient descent update with the initial guess ${}^{S}_{E}q_{k}^{-}$, the recursive solution of \eqref{accmag} can be obtained as
	\begin{equation}
		\begin{aligned}
			{}^{S}_{E}q_{AM,k}={}^{S}_{E}q_{k}^{-}-\mu {\nabla  f_{g,m}}
		\end{aligned}
		\label{gd0}
	\end{equation}
	with 
	\begin{equation}
	{\nabla  f_{g,m}}=\frac{\partial E_k}{\partial {}^{S}_{E}q_{k}}\frac{\partial f_{g,m}}{\partial E_k}=J_{E_k}({}^{S}_{E}q_k)E_k
	\label{gd1}
	\end{equation}
	where $J_{E_k}({}^{S}_{E}q_k)=\frac{\partial E_k}{\partial {}^{S}_{E}q_{k}}=\left[ \frac{\partial E_k}{\partial {}^{S}_{E}q_{k}}\big|_{{}^{S}_{E}q_{k}^{-}}\right]$, ${}^{S}_{E}q_{AM,k}$ is the quaternion obtained by accelerometer and magnetometer readings and $\mu$ is the learning rate. Following the method used in \cite{b7}, we modify \eqref{gd0} as
    \begin{equation}
		\begin{aligned}
			{}^{S}_{E}q_{AM,k}={}^{S}_{E}q_{k}^{-}-\mu \frac{\nabla  f_{g,m}}{\|\nabla  f_{g,m}\|_2}.
			\label{gd}
		\end{aligned}
	\end{equation}
	\begin{remark}
     \color{black}
	The benefit of using normalized gradient direction is to increase the robustness with respect to unexpected large $E_k$, at the price of decreased smoothness when $E_k$ is small (this is the reason why the GD seems much more noisy than the KF-based method when the IMU has slow-varying dynamics). 
	\color{black}
	\end{remark}
	Combining \eqref{gyrq1} and \eqref{gd} in a complementary form gives
	\begin{equation}
	{}^{S}_{E}q_{k} = \gamma {}^{S}_{E}q_{k}^{-} + (1- \gamma) {}^{S}_{E}q_{AM,k}
	\label{fusion}
	\end{equation}
	where $\gamma$ is a coefficient to balance between the gyroscope estimation and the accelerometer and magnetometer estimation. Substituting \eqref{gd} into \eqref{fusion}, one has
	\begin{equation}
	{}^{S}_{E}{q}_{k}={}^{S}_{E}{q}_{k}^{-}-\mu(1-\gamma)\frac{\nabla  f_{g,m}}{\|\nabla  f_{g,m}\|_2}.
	\label{fusionNew}
	\end{equation}
	By setting $\lambda=(1-\gamma)\mu$, one obtains
	\begin{equation}
	\begin{aligned}
	{}^{S}_{E}{q}_{k}={}^{S}_{E}{q}_{k}^{-}-\lambda\frac{\nabla  f_{g,m}}{\|\nabla  f_{g,m}\|_2}
	\end{aligned}
	\label{fusion_new}
	\end{equation}
	where $\beta$ is the only parameter to be tuned in the GD. The whole algorithm is then summarized as follows:
	\begin{itemize}
		\item Prediction: estimate the orientation using \eqref{gyrq1}.
		\item Update: correct the orientation using \eqref{fusion_new}.
	\end{itemize}
	\begin{remark}
	The detailed derivation procedure of the GD described in the above paragraphs is slightly different from the original work~\cite{b7}, but in fact, they are mathematically equivalent.
	\end{remark}
	\subsubsection{The DOE}  
	\label{DOE}
	The DOE algorithm in ~\cite{b14} shares the same strategy in terms of the \emph{a priori} estimate of orientation using gyroscope readings as shown in \eqref{gyrq1}. The difference is that it employs an analytical solution for the orientation correction. Define 
$$
{}^{E}g=\begin{bmatrix}
	0\\
	{}^\mathrm{E}\mathrm{r}_\mathrm{a}
\end{bmatrix}, 
{}^{S}g=\begin{bmatrix}
	0\\
	{}^\mathrm{S}\mathrm{a}_\mathrm{k}
\end{bmatrix},
{}^{E}m=\begin{bmatrix}
	0\\
	{}^\mathrm{E}\mathrm{r}_\mathrm{m}
\end{bmatrix},{}^{S}m=\begin{bmatrix}
	0\\
	{}^\mathrm{S}\mathrm{m}_\mathrm{k} 
\end{bmatrix}$$
where ${}^\mathrm{E}\mathrm{r}_\mathrm{a}=[0,0,1]^{T}$ and ${}^\mathrm{E}\mathrm{r}_\mathrm{m}=[1,0,0]^{T}$ are the reference $z$-axis and $x$-axis in the earth frame, and ${}^\mathrm{S}\mathrm{a}_\mathrm{k}={{y}}_{A,k}$ and ${}^\mathrm{S}\mathrm{m}_\mathrm{k}={{y}}_{M,k}$ are the normalized accelerometer and magnetometer readings. Then, equations \eqref{acc} and \eqref{mag} can be equivalently written as
\begin{subequations}
	\begin{align}
		\begin{aligned}
			\min_{{}^{S}_{E}q_{k} \in \mathbb{R}^{4}} f_g({}^{E}g,{}^{S}g, {}^{S}_{E}q_{k})=\frac{1}{2}{\|}E_{g,k}\|^{2}
			\label{acc1}
		\end{aligned}
	\end{align}
	\begin{align}
		\begin{aligned}
			\min_{{}^{S}_{E}q_{k} \in \mathbb{R}^{4}} f_m({}^{E}m,{}^{S}m,{}^{S}_{E}q_{k})=\frac{1}{2}\|E_{m,k}\|^{2}
			\label{mag1}
		\end{aligned}
	\end{align}
	\label{accmagfun}
\end{subequations}
with 
\begin{subequations}
	\begin{align}
		E_{g,k}&=\Big{\|}{}^{S}_{E}q_{k}^{*}\otimes \begin{bmatrix}
			0\\
			{}^\mathrm{E}\mathrm{r}_\mathrm{a}
		\end{bmatrix}  \otimes{}^{S}_{E}q_{k} - \begin{bmatrix}
			0\\
	        {}^\mathrm{S}\mathrm{a}_\mathrm{k}
		\end{bmatrix} \Big{\|}^{2}
	\end{align}
	\begin{align}
		E_{m,k}&=\Big{\|}{}^{S}_{E}q_{k}^{*}\otimes \begin{bmatrix}
			0\\
			{}^\mathrm{E}\mathrm{r}_\mathrm{m}
		\end{bmatrix} \otimes{}^{S}_{E}q_{k} - \begin{bmatrix}
			0\\
	        {}^\mathrm{S}\mathrm{m}_\mathrm{k}
		\end{bmatrix} \Big{\|}^{2}.
	\end{align}
\end{subequations}
Note that here $[0,{}^\mathrm{E}\mathrm{r}_\mathrm{m}]^{T}$ in \eqref{mag1} is slightly different with ${}^E m$ in \eqref{mag} since the magnetic dip is ignored in \cite{b14}. \textcolor{black}{The above optimization system is overdetermined since each of the two equations provides two independent constraints on ${}^{S}_{E}q_{k}$, whereas ${}^{S}_{E}q_{k}$ only has three degrees of freedom. Conventionally, we can solve \eqref{accmagfun} by minimizing both $E_{g,k}$ and $E_{m,k}$ simultaneously. However, it could result in disturbances in the magnetic field affecting the roll and pitch. To solve this problem, an analytical solution of \eqref{accmagfun} was developed in  \cite{b14} by constraining the magnetometer readings only affecting the heading.} Specifically, for sub-problem \eqref{acc1}, the included angle between ${}^\mathrm{S}\mathrm{r}_\mathrm{a}$ and ${}^\mathrm{S}\mathrm{a}_\mathrm{k}$ has $\alpha_{err,a}=\sphericalangle ({}^\mathrm{S}\mathrm{r}_\mathrm{a}, {}^\mathrm{S}\mathrm{a}_\mathrm{k}) = \arccos(\frac{\langle{}^\mathrm{S}\mathrm{r}_\mathrm{a} ,{}^\mathrm{S}\mathrm{a}_\mathrm{k}\rangle}{\|{}^\mathrm{S}\mathrm{r}_\mathrm{a}\|_2 \|{}^\mathrm{S}\mathrm{a}_\mathrm{k}\|_2})$ where $\langle\cdot ,\cdot \rangle$ is the inner product. Moreover, the rotation axis can be written as $\mathrm{x}_{a}= \frac{  {}^\mathrm{S}\mathrm{r}_\mathrm{a} \times {}^\mathrm{S}\mathrm{a}_\mathrm{k}}{\|{}^\mathrm{S}\mathrm{r}_\mathrm{a} \times {}^\mathrm{S}\mathrm{a}_\mathrm{k}\|_2}$. Finally, the analytical solution of \eqref{acc1} has
\begin{equation}
	\begin{aligned}
		{}^{S}_{E}q_{acc,k}&= \begin{bmatrix}
			\cos(\frac{1}{2}\alpha_{err,a})\\
			\sin(\frac{1}{2}\alpha_{err,a})\mathrm{x}_{a}\\
		\end{bmatrix}
		\label{solutionacc}
	\end{aligned}
\end{equation}
where ${}^{S}_{E}q_{acc,k}$ denotes the quaternion obtained by the accelerometer readings. Since the accelerometer measurement is noisy, it is feasible to use only a small portion of $\alpha_{err,a}$ to correct the gyroscope drift, which gives 
\begin{equation}
	{}^{S}_{E}{q}_{ga,k}={}^{S}_{E}q_{k}^{-}\otimes q_{acor,k}
	\label{acccor}
\end{equation}
with 
\begin{equation}
	q_{acor,k}= \begin{bmatrix}
		\cos(\frac{1}{2}k_{a}\alpha_{err,a})\\
		\sin(\frac{1}{2}k_{a}\alpha_{err,a})\mathrm{x}_{a}\\
	\end{bmatrix}
	\label{qacor}
\end{equation}
where $k_{a} \in (0,1)$ is an adjustable weight, ${}^{S}_{E}q_{k}^{-}$ is the \emph{a priori} estimate of the quaternion as shown in \eqref{gyrq1}, and ${}^{S}_{E}{q}_{ga,k}$ is the obtained quaternion by fusing the gyroscope and accelerometer measurements. 
	
	To fulfill the constraint that magnetometer readings only affect the heading, one can project ${}^\mathrm{S}\mathrm{m}_\mathrm{k}$ into the horizontal plane (in the sensor frame), i.e, 
	${}^\mathrm{S}\bar{\mathrm{m}}_\mathrm{k} =  {}^\mathrm{S}\mathrm{m}_\mathrm{k} - ({}^\mathrm{S}\mathrm{m}_\mathrm{k} \cdot {}^\mathrm{S}\mathrm{r}_\mathrm{a})\cdot   {}^\mathrm{S}\mathrm{r}_\mathrm{a}$. Then, the included angle between ${}^\mathrm{S}\bar{\mathrm{m}}_\mathrm{k}$ and ${}^\mathrm{S}\mathrm{r}_{m}$ has $\alpha_{err,m}= \sphericalangle ({}^\mathrm{S}\bar{\mathrm{m}}_\mathrm{k}, {}^\mathrm{S}\mathrm{r}_\mathrm{m}) = \arccos(\frac{\langle{}^\mathrm{S}\bar{\mathrm{m}}_\mathrm{k},{}^\mathrm{S}\mathrm{r}_\mathrm{m} \rangle}{\|{}^\mathrm{S}\bar{\mathrm{m}}_\mathrm{k}\|_2 \|{}^\mathrm{S}\mathrm{r}_\mathrm{m}\|_2})$ and the rotation axis has $\mathrm{x}_{m}=\frac{{}^\mathrm{S}\bar{\mathrm{m}}_\mathrm{k} \times {}^\mathrm{S}{\mathrm{r}_\mathrm{m}}} { \|{}^\mathrm{S}\bar{\mathrm{m}}_\mathrm{k} \times  {}^\mathrm{S}\mathrm{r}_\mathrm{m} \|_2}$. Finally, by analogy to \eqref{solutionacc} and \eqref{acccor}, one has
	\begin{equation}
	{}^{S}_{E}{q}_{k}= {}^{S}_{E}{q}_{ga,k} \otimes q_{mcor,k}
	\label{magcor}
	\end{equation}
	with
	\begin{equation}
	q_{mcor,k} = \begin{bmatrix}
	\cos(\frac{1}{2}k_{m}\alpha_{err,m})\\
	\sin(\frac{1}{2}k_{m}\alpha_{err,m})\mathrm{x}_{m}\\
	\end{bmatrix}
	\label{qmcor}
	\end{equation}
	where $k_{m} \in (0,1)$ is an adjustable weight and ${}^{S}_{E}{q}_{k}$ is fused quaternion at time step $k$. To further mitigate the gyroscope drift $b_{k}$, the following gyroscope bias update equation is utilized
	\begin{equation}
	b_{k}=b_{k-1}+k_{b,a}\alpha_{err,a}\mathrm{x}_{a}+k_{b,m}\alpha_{err,m}\mathrm{x}_{m}
	\label{gyrbias}
	\end{equation}
	where $k_{b,a},k_{b,m} \in (0,1)$ are two adjustable gains. The bias estimation $b_{k}$ is then utilized to correct the gyroscope readings $\mathrm{y}_{G,k}$ with
	\begin{equation}
	\mathrm{y}_{G,k}^{c}=\mathrm{y}_{G,k}-b_{k}
	\label{drifCorr}
	\end{equation}
	where $\mathrm{y}_{G,k}$ is the raw gyroscope readings and $\mathrm{y}_{G,k}^{c}$ is the corrected one. Correspondingly, $\mathrm{y}_{G,k}^{q}=[0, (\mathrm{y}_{G,k}^{c})^{T}]^{T}$ is utilized in
	\eqref{gyrq1} ro replace $\mathrm{y}_{G,k}^{q}=[0, \mathrm{y}_{G,k}^{T}]^{T}$. The whole algorithm of the DOE is summarized as follows:
	\begin{itemize}
		\item Predict the orientation by gyroscopes using \eqref{gyrq1} and \eqref{drifCorr}.
		\item Update the orientation by accelerometers using \eqref{acccor}.
		\item Update the orientation by magnetometers using \eqref{magcor}. 
	\end{itemize}
	\subsection{The CGD}
	To increase the robustness of the GD with respect to external disturbances, we use the following MKCL-based objective function
	\begin{equation}
	\min_{{}^{S}_{E}q \in \mathbb{R}^{4}} f_{cl}({}^{E}g, {}^{E}m, {}^{S}g, {}^{S}m, {}^{S}_{E}q_{k})=\sum_{i=1}^{8}\sigma_{i}^2\Big(1-G_{\sigma_i}(E_{i,k})\Big)
	\label{accmag1}
	\end{equation}
	where $E_{i,k}$ is $i$-th element of $E_{k}$, $\sigma_{i}$ is $i$-th bandwidth for channel $i$. Correspondingly, the derivative of $f_{cl}$ with respect to ${}^{S}_{E}q_{k}$ can be calculated as
	\begin{equation}
	\begin{aligned}
	\nabla f_{cl} &= \left[\frac{\partial E_{k}}{\partial {}^{S}_{E}q_{k}}\bigg|_{{}^{S}_{E}q_{k}^{-}}\right] \frac{\partial f_{cl}}{\partial E_{k}}= J_{E_k}({}^{S}_{E}q_k) W E_k
	\label{nablaf}
	\end{aligned}
	\end{equation}
	with
	\begin{equation}
	\begin{aligned}
	&W= \operatorname{diag}[G_{\sigma_i}(E_{1,k}),G_{\sigma_i}(E_{2,k})\ldots,G_{\sigma_i}(E_{8,k})]\\
	&E_{k}=\begin{bmatrix}
	E_{g}\\
	E_{m}
	\end{bmatrix}, J_{E_k}({}^{S}_{E}q_{k})=\begin{bmatrix}
	J_{E_g}({}^{S}_{E}q_k), J_{E_m}({}^{S}_{E}q_k)
	\end{bmatrix}\\
	&E_{g}={\begin{bmatrix}
		0\\
		2(q_{2}q_{4}-q_{1}q_{3})-y_{ax}\\
		2(q_{1}q_{2}+q_{3}q_{4})-y_{ay}\\
		2(0.5 -q_{2}^{2}-q_{3}^{2})-y_{az}
		\end{bmatrix}}\\
	&E_{m}={\begin{bmatrix}
		0\\
		2m_x(0.5-q_3^2-q_4^2)+2m_z(q_2q_4-q_1q_3)-y_{mx}\\
		2m_x(q_2q_3-q_1q_4)+2m_z(q_1q_2+q_3q_4)-y_{my}\\
		2m_x(q_2q_3+q_1q_4)+2m_z(0.5-q_2^2-q_3^2)-y_{mz}
		\end{bmatrix}}\\
		&J_{E_g}({}^{S}_{E}q_k)=\begin{bmatrix}
	0&0&0&0\\
	-2q_3 & 2q_4 & -2q_1 & 2q_2\\
	2q_2 & 2 q_1 & 2 q_4 & 2q_3\\
	0 & -4 q_2 & -4 q_3 & 0
	\end{bmatrix}\\
	&J_{E_m}({}^{S}_{E}q_k)=\left[\begin{array}{cc}
	0&0\\
	-2 m_{z} q_{3} & 2 m_{z} q_{4} \\ -2 m_{x} q_{4}+2 m_{z} q_{2} & 2 m_{x} q_{3}+2 m_{z} q_{1} \\ 2 m_{x} q_{3} & 2 m_{x} q_{4}-4 m_{z} q_{2} \end{array}\right.\\
	&\left.\begin{array}{cc}
	0&0\\
	-4 m_{x} q_{3}-2 m_{z} q_{1} & -4 m_{x} q_{4}+2 m_{z} q_{2} \\ 2 m_{x} q_{2}+2 m_{z} q_{4} & -2 m_{x} q_{1}+2 m_{z} q_{3} \\ 2 m_{x} q_{1}-4 m_{z} q_{3} & 2 m_{x} q_{2}\end{array}\right].
\end{aligned}
\label{measurement}
\end{equation}
where ${}^{S}_{E}q_{k}^{-}=[q_1,q_2,q_3,q_4]^{T}$ is the quaternion obtained by gyroscope readings as shown in \eqref{gyrq1}. One can observer that the only difference between \eqref{gd1} and \eqref{nablaf} is the attenuation matrix $W$.
By analogy to equations \eqref{gd}--\eqref{fusion_new}, we have
\begin{equation}
	\begin{aligned}
		{}^{S}_{E}{q}_{k}={}^{S}_{E}{q}_{k}^{-}-\lambda \frac{\nabla f_{cl}}{\|\nabla f_{g,m}\|_2}.
	\end{aligned}
	\label{fusion_new1}
\end{equation}

In a practical application, we can use $\sigma_2=\sigma_3=\sigma_4=\sigma_a$ for the accelerometer readings and $\sigma_6=\sigma_7=\sigma_8=\sigma_m$ for the magnetometer readings due to the homogeneous property of sensors at different axes. The selections of $\sigma_1$ and $\sigma_5$ are redundant since $E_{1,k}=E_{5,k}=0$ always holds (note that the error quaternion is composed of error vectors). \textcolor{black}{A remaining question is to obtain the initial quaternion ${}^{S}_{E}{q}_{0}$. In this work, we use \emph{ecompass} algorithm~\cite{c23} to initialize the orientation. Specifically, the initial rotation matrix is calculated by
\begin{equation}
	R_{0}=\begin{bmatrix}
		({y}_{A}\times {{y}}_{M})\times {y}_{A}&(-{y}_{A})\times {{y}}_{M}& (-{y}_{A})
	\end{bmatrix}
	\label{init}
\end{equation}   
where ${{y}}_{A}$ and ${{y}}_{M}$ are the initial normalized accelerometer and magnetometer readings, and the operator $\times$ denotes the cross product. Then, we convert this rotation matrix to a quaternion and obtain ${}^{S}_{E}{q}_{0}$.} The detailed algorithm of the CGD is summarized in Algorithm \ref{cgd}.
\begin{algorithm}[!t]
   	\caption{The CGD}
   	\label{cgd}
   	\begin{algorithmic}[1]
   		\State {Initialize ${}^{S}_{E}q_{0}$} by \eqref{init}
   		\While{$k>0$}\\
   		Predict the quaternion ${}^{S}_{E}q_{k}^{-}$ using \eqref{gyrq1}\\
   		Obtain the measurement error and the gradient using \eqref{measurement}\\
   		Update the quaternion using \eqref{fusion_new1}\\
   		$k=k+1$
   		\EndWhile 
   	\end{algorithmic}
\end{algorithm}
	\subsection{The CDOE}
	Based on the algorithm of the DOE described in Section \ref{DOE}, we can remodel the correction angles $\alpha_{err,a}$ and $\alpha_{err,m}$ as
	\begin{equation}
		\begin{aligned}
		\alpha_{err,a}&= \alpha_{ca} + v_{ca}\\
		\alpha_{err,m}&= \alpha_{cm} + v_{cm}\\
		\end{aligned}
	\end{equation}
	where $\alpha_{ca}$ and $\alpha_{cm}$ is the ground truth correction angle we want to obtain, and $v_{ca}$ and $v_{cm}$ are the noises caused by accelerometer and magnetometer measurement noises. By assuming that $v_{ca}$ and $v_{cm}$ are Gaussian, the obtainment of $k_a\alpha_{err,a}$ and $k_m\alpha_{err,m}$ in \eqref{qacor} and \eqref{qmcor} can be given by solving the following LS-based optimization problems:
	\begin{subequations}
	\begin{align}
	\min_{\alpha_{ca} \in \mathbb{R}} f_{a}(\alpha_{ca})= \frac{1}{2}(\alpha_{ca}-\alpha_{err,a})^2,
	\label{alphaAcc}
	\end{align}
	\begin{align}
	\min_{\alpha_{cm} \in \mathbb{R}} f_{m}(\alpha_{cm})= \frac{1}{2}(\alpha_{cm}-\alpha_{err,m})^2.
	\label{alphaMag}
	\end{align}
	\end{subequations}
	Solving \eqref{alphaAcc} and \eqref{alphaMag} with the gradient descent strategy, we have
	\begin{subequations}
	\begin{align}
	\begin{aligned}
	\alpha_{ca,t} &=\alpha_{ca,t-1} - k_{a} \nabla f_a
	\label{cagd}
	\end{aligned}
	\end{align}
	\begin{align}
	\begin{aligned}
	\alpha_{cm,t}&=\alpha_{ca,t-1} - k_{a} \nabla f_m
	 \label{cmgd}
	\end{aligned}
	\end{align}
	\end{subequations}
	where $\nabla f_a= \frac{\partial f_a}{\partial \alpha_{ca}}\big|_{\alpha_{ca,t-1}}=\alpha_{ca,t-1}-\alpha_{err,a}$, $\nabla f_m=\frac{\partial f_m}{\partial \alpha_{cm}}\big|_{\alpha_{cm,t-1}}= \alpha_{ca,t-1}-\alpha_{err,m}$, $t$ is the iteration number and starts from 1, $\alpha_{ca,t-1}$ and $\alpha_{cm,t-1}$ are the correction angles at the previous iteration, and $k_a$ and $k_m$ are step sizes. Note that the initial guess of the correction angle should be zero (since the \emph{a priori} estimate of the correction angle is zero) and it is possible to iterate only one time at each time interval. Thus, we have $\alpha_{ca}=k_{a}\alpha_{err,a}$ and $\alpha_{cm}=k_{m}\alpha_{err,m}$ (ignoring the subscript $t$), which are identical to the corrected angles utilized in quaternion update in \eqref{qacor} and \eqref{qmcor}.
	
	Since the Gaussian assumption of $v_{ca}$ and $v_{cm}$ is generally not valid due to the involvement of external acceleration and magnetic disturbance, we construct the following MKCL-based objective functions:
	\begin{subequations}
	\begin{align}
	\min_{{\alpha}_{ca} \in \mathbb{R}} f_{cl,a}({\alpha}_{ca} )= \sigma_{{a}}^2 \big(1-G_{\sigma_{{a}}}(e_{a})\big),
	\end{align}
	\begin{align}
	\min_{{\alpha}_{cm} \in \mathbb{R}} f_{cl,m}({\alpha}_{cm})= \sigma_{{m}}^2 \big(1-G_{\sigma_{{m}}}(e_{m})\big)
	\end{align}
	\end{subequations}
	where $e_{a}={\alpha}_{ca}-\alpha_{err,a}$, $e_{m}={\alpha}_{cm}-\alpha_{err,m}$, and $\sigma_{{a}}$ and $\sigma_{{m}}$ are two bandwidths. By analogy to the one-step gradient descent update in \eqref{cagd} and \eqref{cmgd}, we have
	\begin{equation}
	\begin{aligned}
	{\alpha}_{ca}&=k_a G_{\sigma_{{a}}}(\alpha_{err,a})\alpha_{err,a}\\
	{\alpha}_{cm}&=k_m G_{\sigma_{{m}}}(\alpha_{err,m})\alpha_{err,m}
	\end{aligned}
	\label{corangle}
	\end{equation} and thus
	\begin{equation}
	{q}_{acor,k}= \begin{bmatrix}
	\cos(\frac{1}{2}{\alpha}_{ca})\\
	\sin(\frac{1}{2}{\alpha}_{ca})\mathrm{x}_{a}\\
	\end{bmatrix}, 
	{q}_{mcor,k}= \begin{bmatrix}
	\cos(\frac{1}{2}{\alpha}_{cm})\\
	\sin(\frac{1}{2}{\alpha}_{cm})\mathrm{x}_{m}\\
	\end{bmatrix}.
	\label{errquat}
	\end{equation}
	Similar to \eqref{gyrbias}, the gyroscope drift $b_{k}$ can be updated as
	\begin{equation}
	\begin{aligned}
	b_{k}&=b_{k-1}+k_{b,a}\alpha_{ca} \mathrm{x}_{a}+k_{b,m}\alpha_{cm}\mathrm{x}_{m}.
	\label{drif}
	\end{aligned}
	\end{equation}
	Then, the quaternion update by accelerometer readings is given as 
	\begin{equation}
	{}^{S}_{E}{q}_{ga,k}={}^{S}_{E}q_{k}^{-}\otimes q_{acor,k}.
	\label{acccor1}
	\end{equation}
	Finally, the quaternion update by magnetometer readings is given as 
	\begin{equation}
	{}^{S}_{E}{q}_{k}= {}^{S}_{E}{q}_{ga,k} \otimes q_{mcor,k}.
	\label{magcor1}
	\end{equation}
	The detailed algorithm of the CDOE is summarized in Algorithm \ref{cdoe}.
	\begin{algorithm}[t]
		\caption{The CDOE}
		\label{cdoe}
		\begin{algorithmic}[1]
			\State {Initialize ${}^{S}_{E}q_{0}$ using \eqref{init} and set $b_{0}=0$}
			\While{$k>0$}\\
			Calculate ${}^{S}_{E}q_{k}^{-}$ using \eqref{gyrq1} and \eqref{drifCorr}\\
			Obtain the correction angles $\alpha_{err,a}$ and $\alpha_{err,m}$,  and calculate ${q}_{acor,k}$ and ${q}_{mcor,k}$ using \eqref{corangle} and \eqref{errquat}\\
			Correction by accelerometer readings using \eqref{acccor1}\\
			Correction by magnetometer readings using \eqref{magcor1}\\
			Gyroscope bias estimation using \eqref{drif}\\
			$k=k+1$
			\EndWhile 
		\end{algorithmic}
	\end{algorithm}
\color{black}
\subsection{Kernel Bandwidth Tuning}
\label{tuning}
The kernel bandwidths play important roles in the CGD and CDOE. As indicated by Theorem \ref{theorem1}, the performance of the CGD and CDOE is almost identical to the GD and DOE when using large kernel sizes. In this case, they perform well under Gaussian noise but are sensitive to external disturbance. On the contrary, a very small kernel bandwidth can reject disturbance effectively at the price of degraded ability in inhibiting gyroscope drift. One method for this trade-off is to minimize the ``distance" between the objective function-induced pdf and the practical noise distribution, which falls into the category of MLE. Another method is to treat it as a black-box problem and solve it with some black-box optimization algorithms, e.g., Bayesian optimization in \cite{b16}. Although the above two methods are effective, their performance largely depends on the training set, and obvious generalization errors may occur when the testing situation is different from the training case.   

Inspired by the redescending property of the MKCL as shown in Fig. \ref{inf}, one can select kernel bandwidths manually by inspecting the distribution of measurement residuals. Specifically, the influence function of the MKCL in one dimension case has $\nabla J_{GL}(e)= \frac{J_{CL}(e)}{\partial e}= e \exp^{-\frac{e^2}{2\sigma^2}}$, which gives its maximum (and minimum) value at point $e=\sigma$ (and $e=-\sigma$). In the case of $e=\pm3\sigma$, one has $\nabla J_{GL}(e)\approx \pm 0.011e$, which indicates that the residuals with $|e| \ge 3\sigma$ nearly have no influence on the objective function and hence the residuals polluted by disturbances are expected to locate outside of  the region $[-3\sigma,3\sigma]$. On the contrary, the influence functions of the MKCL and the LS are similar within the region $[-\sigma,\sigma]$, and therefore the residuals under the disturbance-free experiments are expected to be in this region. By these two principles, one can adjust the kernel bandwidth effectively. Empirically, we advise using $\sigma_a=2d_a$  and $\sigma_m=2d_m$ where $d_a$ and $d_m$ are the standard deviations of the accelerometer residuals and magnetometer residuals under the disturbance-free experiments.
  \color{black}
\section{Experiments}
In this section, we compare the CGD and CDOE with the GD~\cite{b7}, DOE~\cite{b14}, ESKF~\cite{b6}, and MKMC~\cite{b16}. We first investigate their performances using a commercial IMU. Then, we implement them on a self-designed low-cost IMU and compare their accuracy under some walking experiments.
\subsection{Performance Validation Using a Commercial IMU}
\begin{figure}[htbp]
	\centering
	\subfigure[Translation]{
		\begin{minipage}[t]{0.5\linewidth}
			\centering
			\includegraphics[width=0.7\columnwidth]{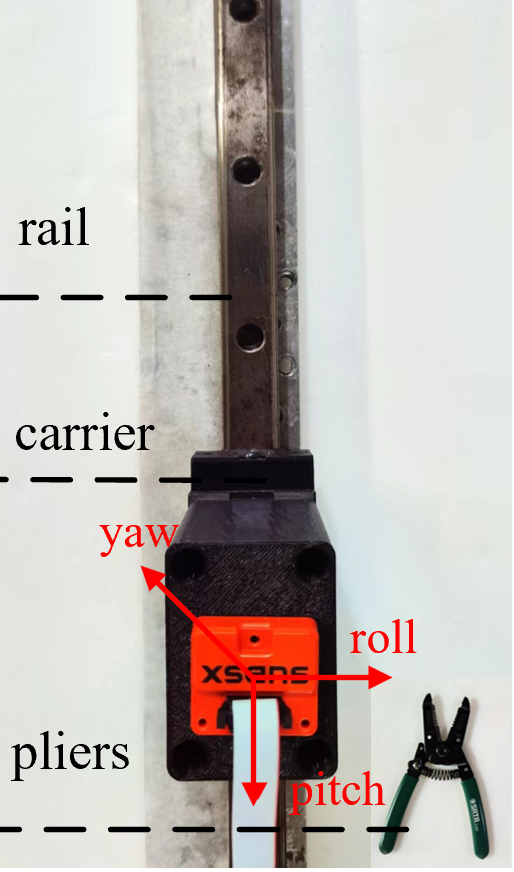}
			\label{linear}	
		\end{minipage}%
	}%
	\subfigure[Translation and rotation]{
		\begin{minipage}[t]{0.525\linewidth}
			\centering
			\includegraphics[width=0.7\columnwidth]{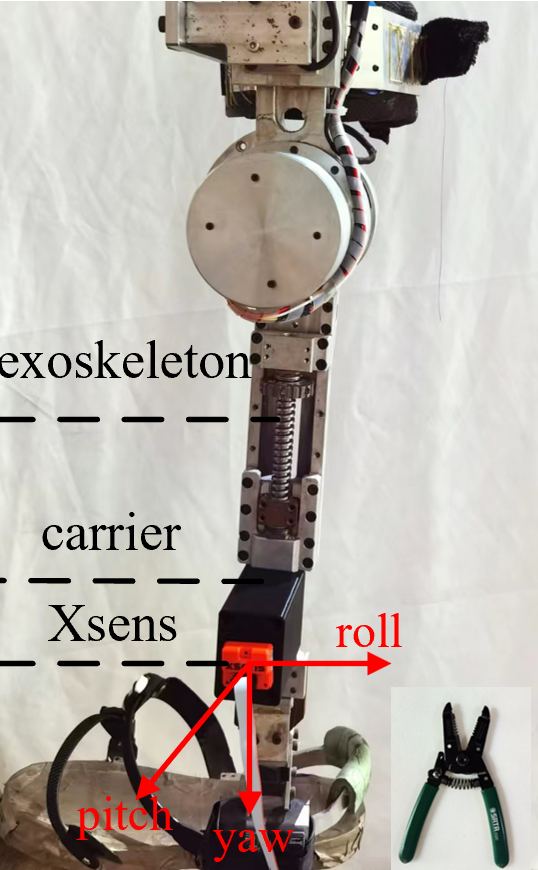}
			\label{rotation}	
		\end{minipage}%
	}%
	\caption{The experimental setups. In Fig. \ref{linear}, Xsens is attached to a carrier that is connected to a rail. The movement of Xsens is controlled manually. The external acceleration $A_d$ is generated by the movement of Xsens, while the magnetic disturbance $M_d$ is caused by the approaching of pliers. In Fig. \ref{rotation}, Xsens is attached to the shank of an exoskeleton, and the exoskeleton is commanded to imitate the walking of a human. The magnitude of $A_d$ depends on the rotation frequency $f$, while $M_d$ is determined by the distance between the pliers and Xsens. \textcolor{black}{The plastic carrier has a height of 12 cm which is sufficient to isolate the magnetic effects caused by rails or exoskeletons.}}		
	\label{walking}
\end{figure}

We validate the performance of different algorithms on a commercial sensor, Xsens MTI-670. The sampling frequency is 400 Hz. Two kinds of motions, translation (T) and translation adjoint with rotation (T \& R), are considered. The experimental setups are shown in Fig. \ref{linear} and Fig. \ref{rotation}, and the detailed experimental descriptions are summarized in Table. \ref{expDesp}. It is worth mentioning that the IMU is disturbed either by acceleration disturbance $A_d$ or magnetic disturbance $M_d$ in experiment 1, while it is disturbed by $A_d$ and $M_d$ simultaneously in experiment 5. Moreover, there is almost \emph{no external disturbance} in experiment 2 since the rotation frequency is very low.
\begin{table}[htbp]
	\setlength{\abovecaptionskip}{1pt}
	\setlength{\belowcaptionskip}{1pt}
	\caption{Experiments Description.}
	\begin{center}
		\scalebox{0.98}{
			\begin{tabular}{ccc}
				\hline
				\hline
				{Exp}& {Movement}&{Disturbance}\\
				\hline
				1&{T}&{$A_d$ or $M_d$}\\
				2&{T \& R}&{$f=0.1Hz$, almost no disturbance}\\
				3&{T \& R}&{$f=0.5Hz$, $A_d$}\\
				4&{T \& R}&{$f=0.1Hz$, $M_d$}\\
				5&{T \& R}&{$f=0.5Hz$, $A_d$ and $M_d$}\\
				\hline
				\hline
		\end{tabular}}
		\label{expDesp}
	\end{center}
\end{table}

The kernel bandwidths of the correntropy-based algorithms are tuned based on the strategies described in Section \ref{tuning}. As for the other parameters, we use the same values for each comparison pair (i.e., the ESKF and MKMC, the GD and CGD, and the DOE and CDOE). The root-mean-square errors (RMSEs) and maximum errors (MEs) of different algorithms under experiments 1-5 are summarized in Table \ref{linearRota}. One can see that the performances of the correntropy-based algorithms (i.e., MKMC, CGD, and CDOE) are slightly better than their counterparts (i.e., ESKF, GD, and DOE) in experiment 2, and significantly outperforms the ESKF, GD, and DOE in other experiments. This reveals that the correntropy-based algorithms can increase the robustness of the traditional algorithms against disturbances but almost without sacrificing their performances under Gaussian noises.	This feature gives its root in the redescending property of the MKCL's influence function.
\begin{table*}[htbp]
	\centering
	\caption{\textcolor{black}{RMSEs and MEs of Different Algorithms Using Xsens.}}
	\scalebox{0.99}{
		\begin{tabular}{cccccccccccccc}
			\hline
			\hline
			\multirow{2}{*}{Test} & \multirow{2}{*}{Axis} & \multicolumn{6}{c}{RMSE ($\deg$)}            & \multicolumn{5}{c}{ME ($\deg$)}       &      \\
			&                       & GD & CGD & DOE & CDOE & {ESKF} & {MKMC} & GD & CGD & DOE & CDOE & {ESKF} & {MKCM} \\
			\hline
			\multirow{3}{*}{1}     & yaw                   & 10.22 & 0.69 & 8.10 & 0.13 & 13.54 & 0.31 & 28.40 & 1.68 & 15.83 & 0.26 & 32.95 & 0.97 \\
			& roll                  & 0.21 & 0.09 & 0.22 & 0.06 & 0.39 & 0.05 & 0.93 & 0.42 & 1.17 & 0.33 & 1.44 & 0.16 \\
			& pitch                 & 0.62 & 0.04 & 0.46 & 0.01 & 0.77 & 0.05 & 3.07 & 0.16 & 2.41 & 0.13 & 4.40 & 0.17\\
			\hline
			\multirow{3}{*}{2}     & yaw                  & 1.87 & 1.84 & 1.29 & 0.95 & 1.34 & 0.50 & 5.55 & 5.41 & 2.88 & 2.16 & 3.82 & 0.82  \\
			& roll           & 0.60 & 0.52 & 0.39 & 0.21 & 0.59 & 0.30 
			& 2.02 & 1.68 & 1.41 & 0.67 & 1.46 & 0.79  \\
			& pitch                & 0.16 & 0.27 & 0.10 & 0.09 & 0.51 & 0.13 & 0.47 & 0.88 & 0.27 & 0.29 & 1.38 & 0.58  \\	
			\hline
			\multirow{3}{*}{3}     & yaw      & 0.94 & 0.44 & 0.74 & 0.53 & 0.67 & 0.15 & 2.69 & 1.29 & 2.65 & 1.18 & 2.01 & 0.50   \\
			& roll                 & 0.79 & 0.11 & 1.88 & 0.23 & 0.82 & 0.14 & 3.26 & 0.65 & 7.72 & 1.11 & 3.35 & 0.79  \\
			& pitch              & 0.23 & 0.16 & 0.44 & 0.14 & 0.33 & 0.14  & 0.82 & 0.77 & 1.40 & 0.73 & 1.47 & 0.58   \\		
			\hline
			\multirow{3}{*}{4}     & yaw                & 17.03 & 2.38 & 12.96 & 2.31 & 17.77 & 0.69 & 31.16 & 4.74 & 27.56 & 3.72 & 34.25 & 1.47\\
			& roll                & 1.67 & 0.40 & 0.34 & 0.17 & 3.48 & 0.23 & 4.94 & 1.35 & 1.20 & 0.58 & 9.93 & 0.69  \\
			& pitch & 1.46 & 0.17 & 0.06 & 0.08 & 2.56 & 0.09 & 4.31 & 0.79 & 0.22 & 0.30 & 6.32 & 0.64  \\
			\hline
			\multirow{3}{*}{5}     & yaw      & 16.30 & 1.29 & 13.12 & 1.67 & 16.60 & 1.42 & 26.63 & 2.64 & 24.05 & 2.93 & 28.81 & 2.98  \\
			& roll             & 1.07 & 0.78 & 1.61 & 0.20 & 2.70 & 0.33  & 3.79 & 2.34 & 7.34 & 1.05 & 8.37 & 1.19\\
			& pitch             & 0.92 & 0.16 & 0.34 & 0.19 & 1.82 & 0.19 & 3.46 & 0.66 & 1.30 & 0.80 & 4.87 & 0.52 \\				
			\hline
			\hline
	\end{tabular}}
	\label{linearRota}
\end{table*}

\begin{figure}[htbp]
	\centerline{\includegraphics[width=8.2cm]{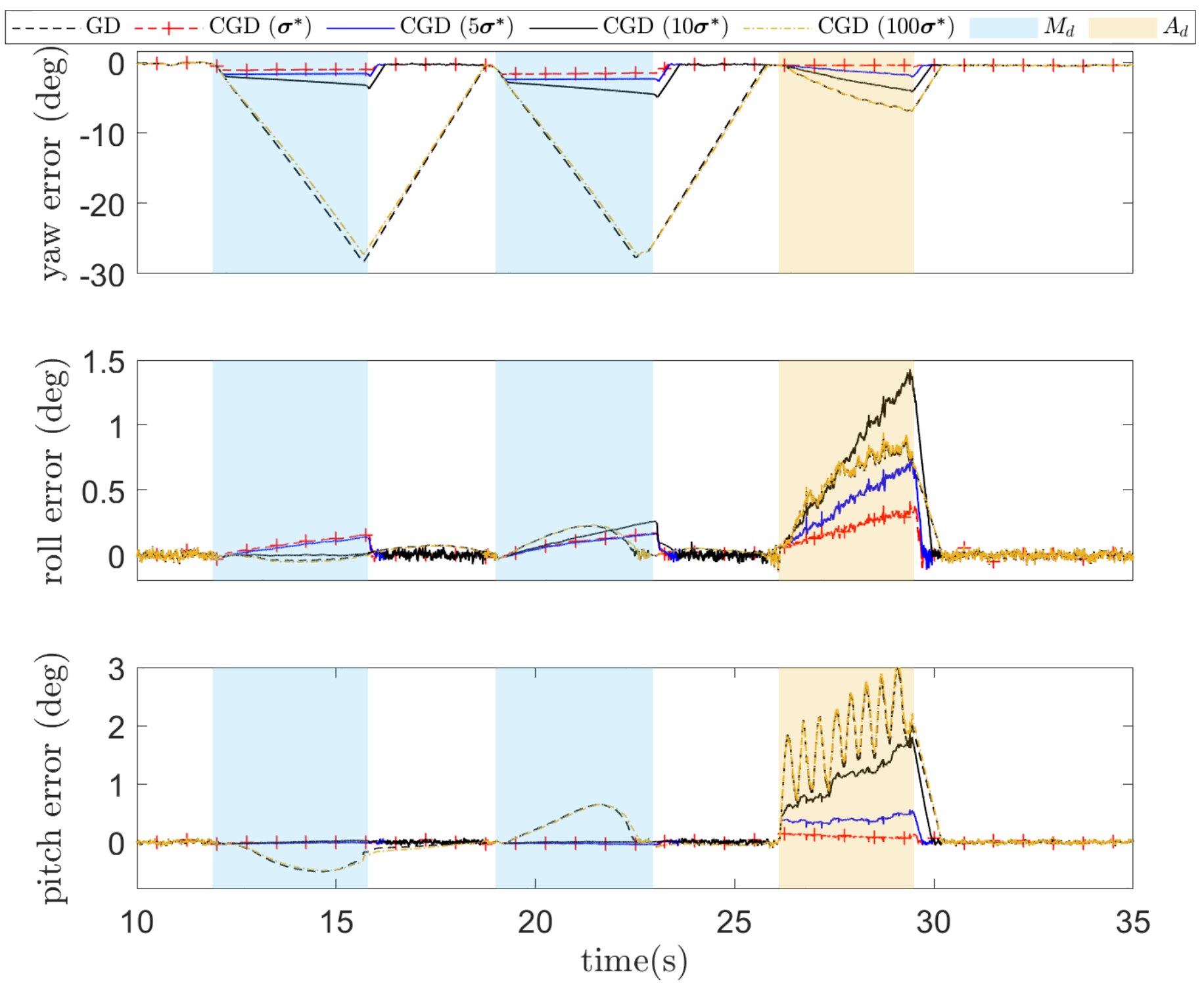}}
	\caption{\textcolor{black}{Performances of the CGD with different kernel bandwidth vectors in experiment 1. The tuned bandwidth vector for the CGD has $\bm{\sigma^{*}}=[\sigma_a,\sigma_m]^{T}=[0.02,0.01]^{T}$.}}
	\label{exp1_gd_band}
\end{figure}
\begin{figure}[htbp]
	\centerline{\includegraphics[width=8.2cm]{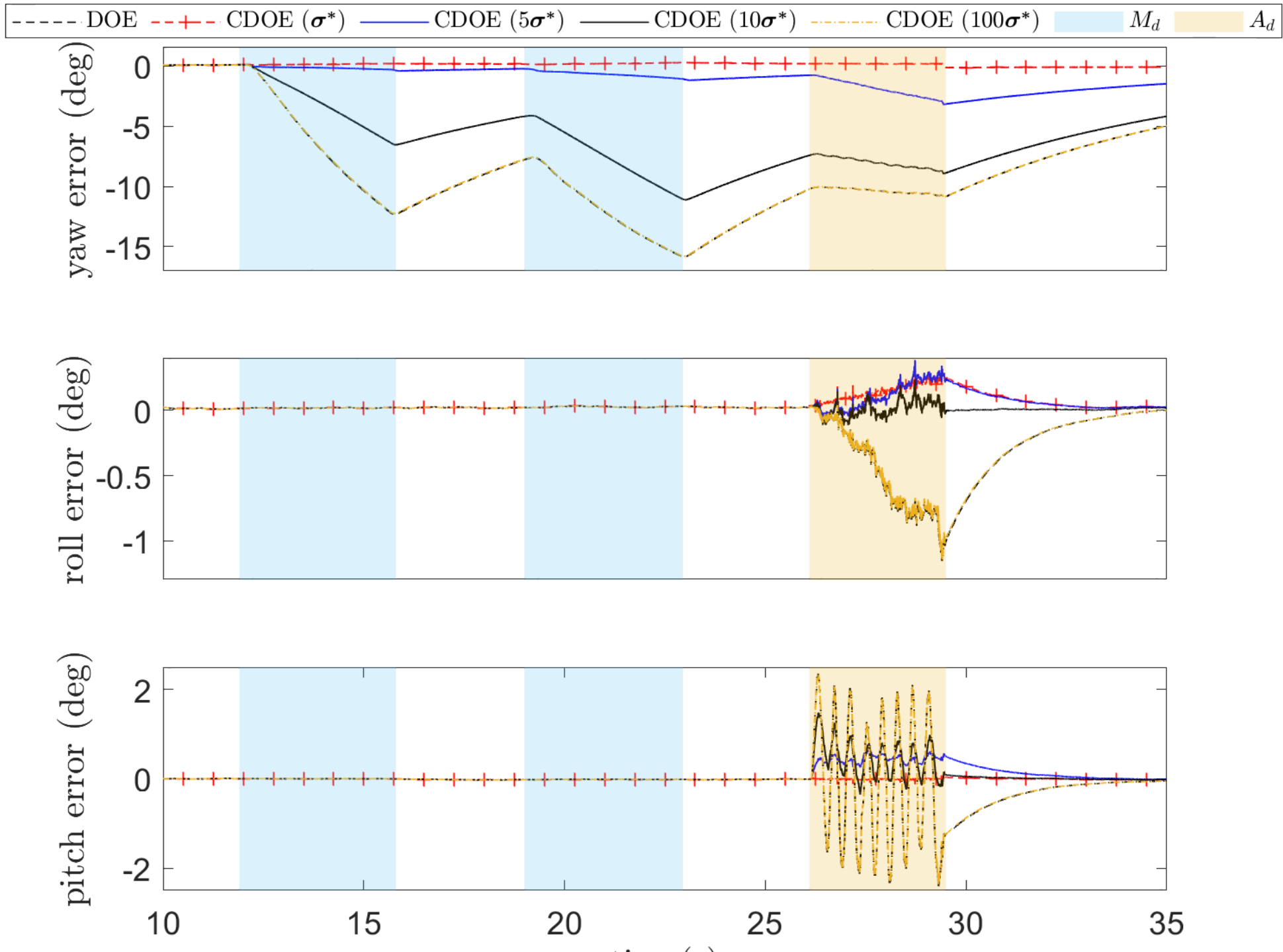}}
	\caption{\textcolor{black}{Performances of the CDOE with different kernel bandwidth vectors in experiment 1. The tuned bandwidth vector for the CDOE has $\bm{\sigma^{*}}=[\sigma_a,\sigma_m]^{T}=[0.05,0.04]^{T}$.}}
	\label{exp1_doe_band}
\end{figure}

\begin{figure*}[htbp]
	\centering
	\subfigure[Norm of sensor readings in experiment 1]{\includegraphics[width=2.78in]{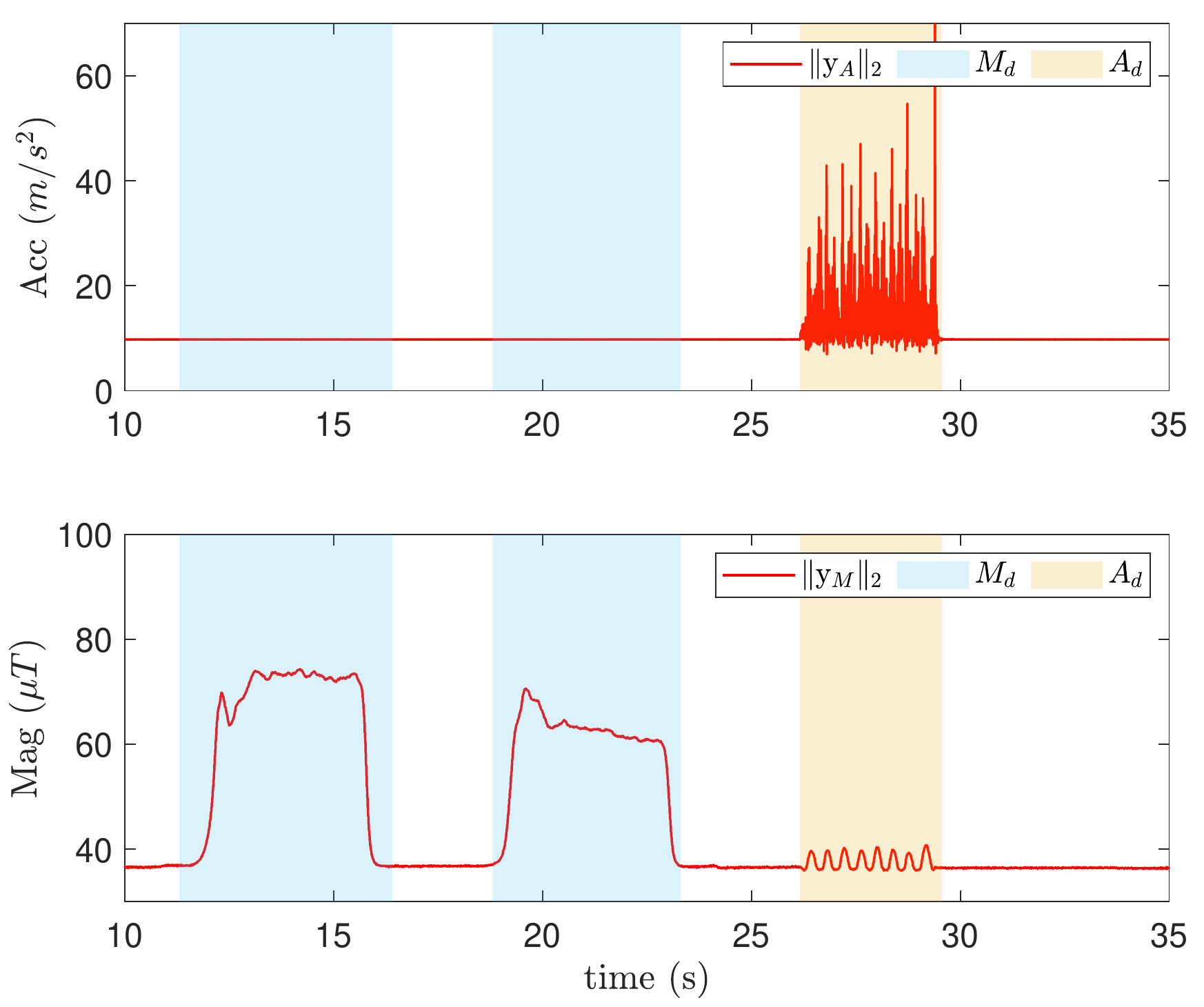}%
		\label{exp1_norm}}
	\subfigure[Orientation errors of different methods in experiment 1]{\includegraphics[width=2.82in]{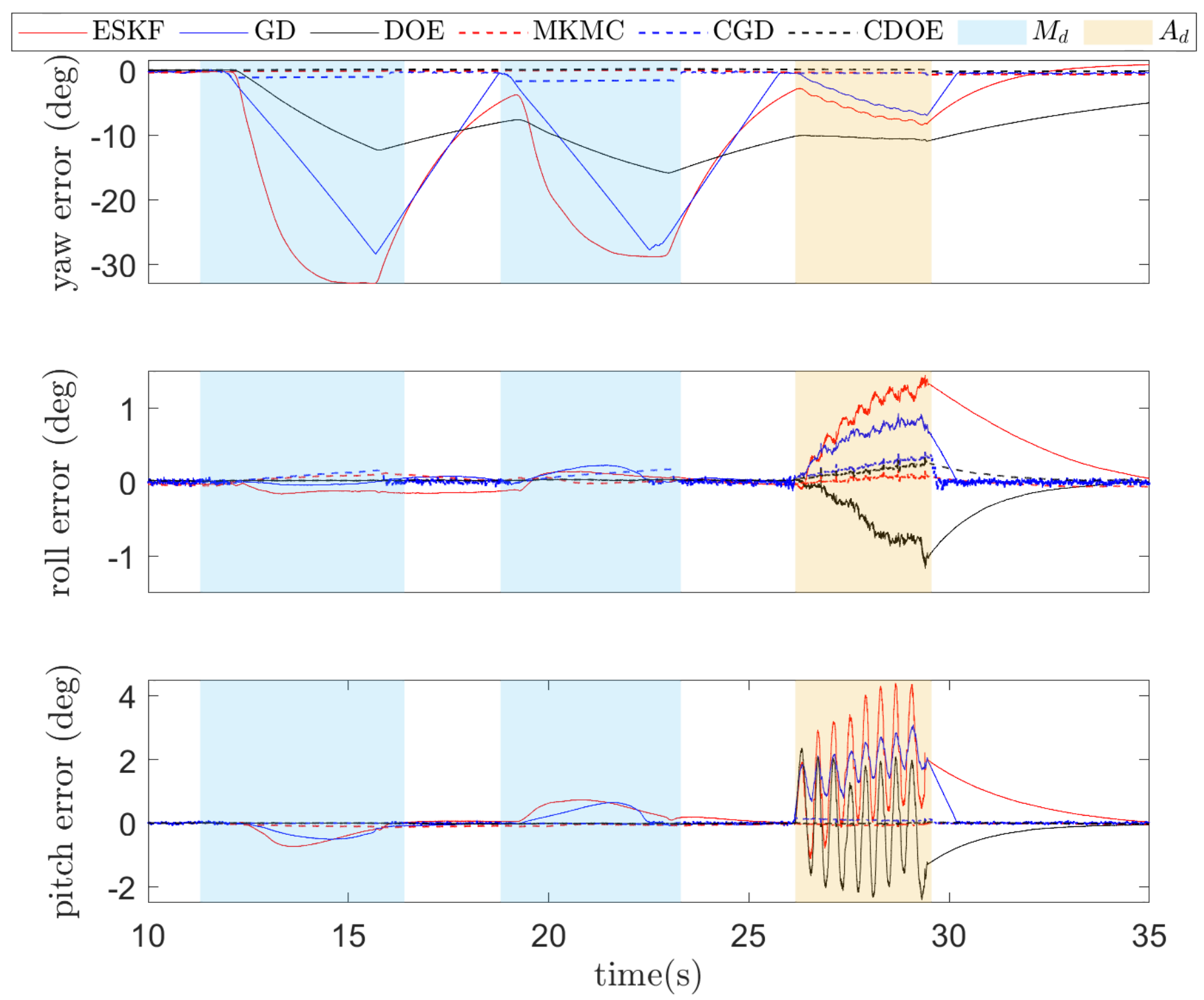}%
		\label{exp1}}
		\\
	\subfigure[Norm of sensor readings in experiment 5]{\includegraphics[width=2.8in]{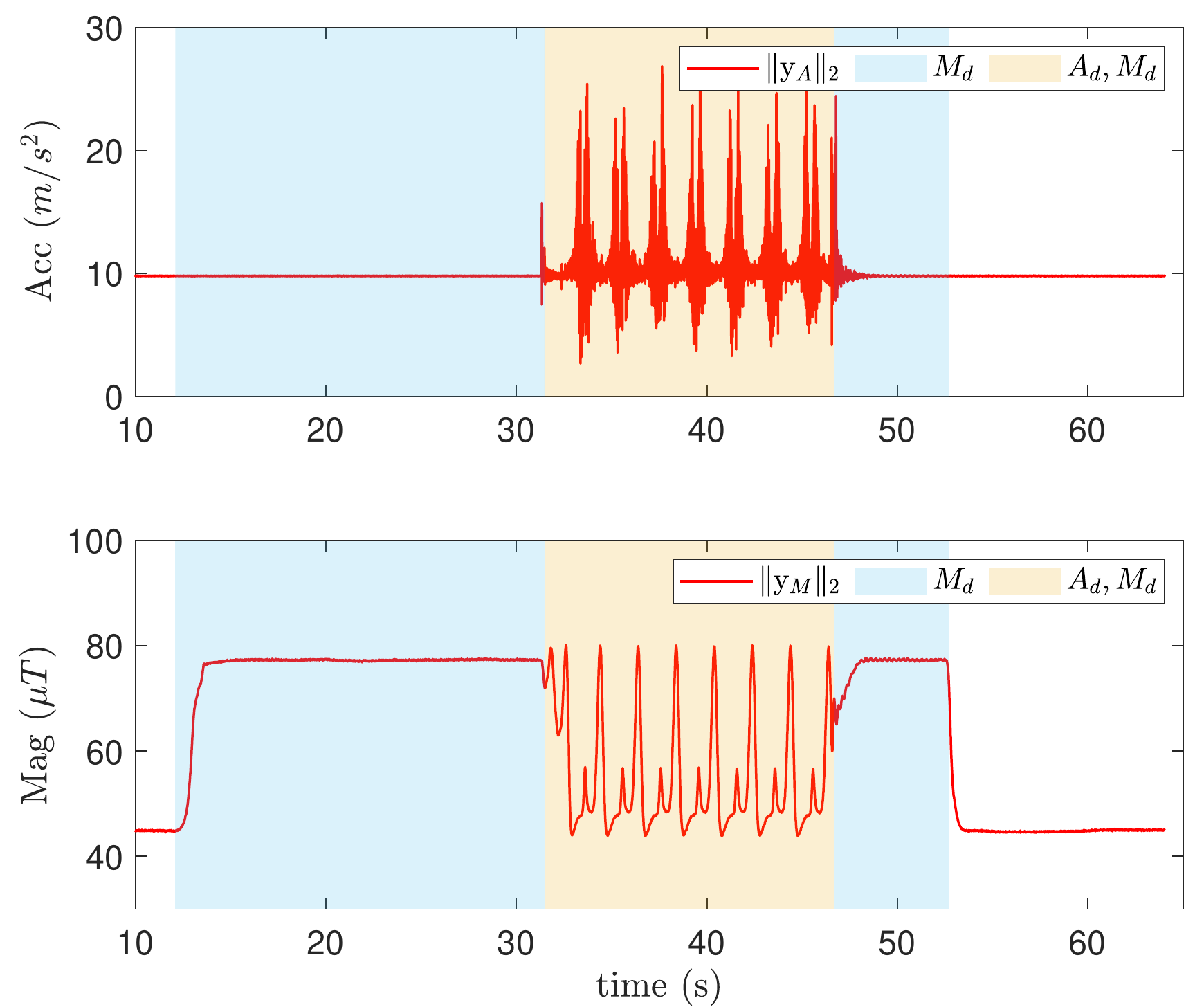}%
	\label{exp5_norm}}
	\subfigure[Orientation errors of different methods in experiment 5]{\includegraphics[width=2.8in]{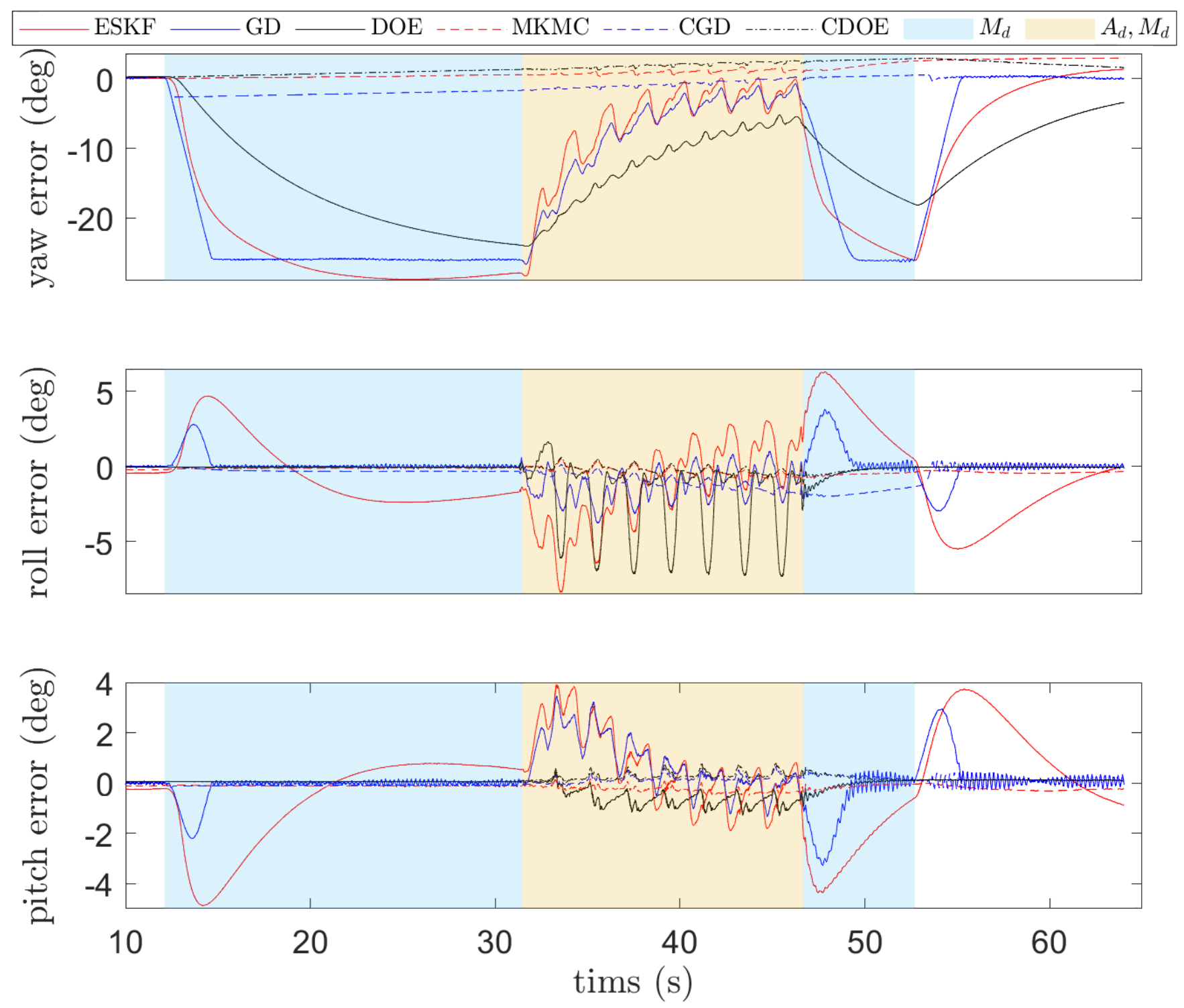}%
	\label{exp5}}
	\caption{\textcolor{black}{Performances of different algorithms in experiments 1 and 5. The left two figures show the norm of the accelerometer and magnetometer readings while the right ones show the orientation errors of different algorithms.}}	
	\label{exp_result}
\end{figure*}

We visualize the error performances of different methods in experiments 1 and 5 in Fig. \ref{exp_result}. The norm of the sensor readings is shown in Figs. \ref{exp1_norm} and \ref{exp5_norm}  while the corresponding errors are shown in Figs. \ref{exp1} and \ref{exp5}. One can observe that three correntropy-based algorithms (i.e., the MKMC, CGD, and CDOE) are very robust to external disturbances, especially along the yaw and pitch axes since these two axes are affected by $A_d$ and $M_d$ the most.

\textcolor{black}{To further investigate the effects of kernel bandwidths on the proposed algorithms, we visualize the orientation errors of the CGD and CDOE by applying different bandwidth vectors in experiment 1 and the corresponding results are shown in Figs. \ref{exp1_gd_band} and \ref{exp1_doe_band}. Not surprisingly, the CGD and CDOE are almost identical to the GD and DOE when applying big kernel bandwidths (see the dashed black and yellow lines), and are very robust to disturbances when using relatively small kernel bandwidths. This result is consistent with the log-likelihood comparison as shown in Fig. \ref{likelihood} which indicates that small kernel bandwidths generally are robust to heavy-tailed noises.}
\subsection{Performance Validation on a Low-cost IMU}
\begin{figure}[htbp]
 		\centerline{\includegraphics[width=8.5cm]{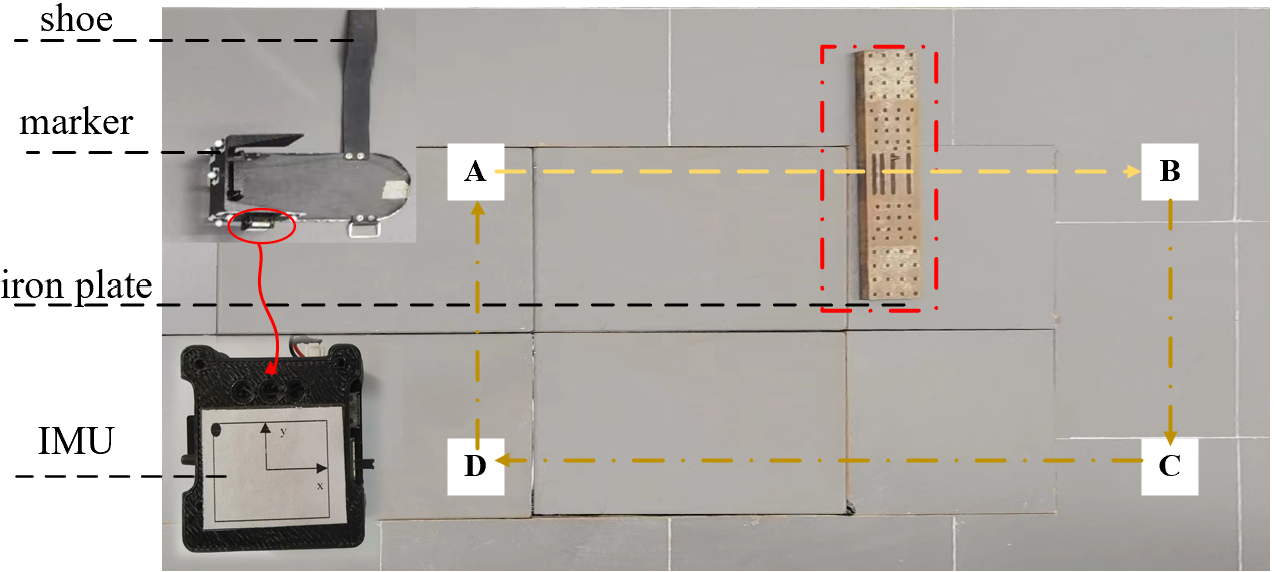}}
 		\caption{Experimental setup for the walking test. The low-cost IMU is attached to a shoe. The external acceleration is generated by the moving of the leg while the magnetic disturbance is generated by the iron plate.}
 		\label{icmWalk}
 	\end{figure}

 	We implement our algorithms to a low-cost IMU (microprocessor: STM32F405RGT6, embedded sensor: ICM20948) which integrates an SD card and an onboard battery. \textcolor{black}{The IMU has been calibrated based on the procedure introduced in ~\cite{c26}.} We attach the IMU to a shoe and investigate its performance under some walking tests. The experimental setup is shown in Fig. \ref{icmWalk} and a detailed experimental description is shown in Table \ref{icm_walk}. \textcolor{black}{The ground truth orientation is obtained by the Vicon motion capture system with 8 cameras (Vicon Vero v2.2), which is highly reliable with 0.017 mm (mean) position tracking error~\cite{c24}. For comparison, the optical data and the IMU output are time-synchronized and aligned. The time synchronization is achieved by correlating the norms of the gyroscope readings and of the angular velocity estimated by the optical system. The alignment is done by an optimization-based method shown in \cite{c25}.} Four experiments are conducted to cover the situations with or without magnetic disturbance and straight walking or walking with turns. 
 	
 	\begin{table*}[]
 		\centering
 		\setlength{\abovecaptionskip}{1pt}
 		\setlength{\belowcaptionskip}{1pt}
 		\caption{\textcolor{black}{RMSEs and MEs of Different Algorithms in the Walking Test.}}
 		\scalebox{0.95}{
 			\begin{tabular}{cccccccccccccc}
 				\hline
 				\hline
 				\multirow{2}{*}{Test} & \multirow{2}{*}{Axis} & \multicolumn{6}{c}{RMSE ($\deg$)}            & \multicolumn{5}{c}{ME ($\deg$)}       &      \\
 				&                       & GD & CGD & DOE & CDOE & ESKF& MKMC & GD & CGD & DOE & CDOE & ESKF & MKMC \\
 				\hline
 				\multirow{3}{*}{1}  &yaw  & 1.00 & 0.95 & 1.02 & 0.46 & 1.20 & 0.70& 4.06 & 2.98 & 6.06 & 1.80 & 7.64 & 2.78   \\
 				& roll             & 1.83 & 0.60 & 4.34 & 0.84 & 2.31 & 0.95 & 5.03 & 1.76 & 16.95 & 2.42 & 7.98 & 2.54 
 				\\
 				& pitch        & 0.75 & 0.87 & 0.98 & 0.74 & 1.02 & 0.91 & 2.59 & 3.30 & 5.48 & 2.56 & 5.23 & 3.24     \\
 				\hline
 				\multirow{3}{*}{2}     & yaw    & 4.78 & 1.28 & 7.90 & 0.69 & 1.84 & 0.64 & 16.08 & 3.08 & 13.23 & 2.10 & 13.38 & 2.49  \\
 				& roll      & 3.15 & 0.33 & 4.24 & 0.47 & 2.69 & 0.34 
 				& 10.35 & 1.14 & 14.34 & 1.47 & 10.45 & 1.23    \\
 				& pitch     & 3.12 & 1.01 & 1.50 & 0.73 & 1.78 & 0.81 
 				 & 10.54 & 2.97 & 8.92 & 2.42 & 11.37 & 2.62  \\  
 				\hline
 				\multirow{3}{*}{3}     & yaw & 6.35 & 2.05 & 2.86 & 1.30 & 2.82 & 1.69 & 17.78 & 9.01 & 8.29 & 3.81 & 16.35 & 5.30    \\
 				& roll   & 1.30 & 0.64 & 3.67 & 0.63 & 2.09 & 0.74 & 4.68 & 2.52 & 14.94 & 2.71 & 10.29 & 2.47  \\
 				& pitch      & 1.42 & 1.29 & 1.92 & 1.07 & 2.79 & 1.61 
 				& 5.43 & 5.85 & 9.24 & 4.57 & 13.74 & 7.06  \\  
 				\hline
 				\multirow{3}{*}{4}     & yaw    & 6.52 & 2.40 & 6.32 & 1.88 & 6.31 & 2.48 & 14.18 & 7.36 & 13.28 & 4.16 & 22.41 & 5.69 \\
 				& roll  & 1.35 & 0.84 & 3.59 & 0.73 & 1.78 & 0.78& 4.85 & 2.45 & 16.36 & 2.28 & 7.68 & 3.11  \\
 				& pitch    & 1.26 & 1.10 & 1.47 & 0.91 & 2.81 & 1.26 & 7.45 & 4.48 & 11.31 & 3.98 & 20.06 & 6.00  \\  
 				\hline
 				\hline
 		\end{tabular}}
 		\label{icmwalkingTest}
 	\end{table*}
  	\begin{table}[htbp]
 	\centering
 	\caption{Experimental Description.}
 	\scalebox{0.9}{
 	\begin{tabular}{lll}
 			\hline
 			\hline
 			Test       & {Route}  & Disturbance                                                    \\
 			1   & A$\rightarrow$B&  $A_d$, without iron plate \\
 			2   & A$\rightarrow$B & $A_d$, $M_d$, with iron plate \\
 			3   & A$\rightarrow$B$\rightarrow$C$\rightarrow$D$\rightarrow$A& $A_d$, without iron plate                        \\
 			4  &A$\rightarrow$B$\rightarrow$C$\rightarrow$D$\rightarrow$A& $A_d$, $M_d$, with iron plate   \\
 			\hline
 			\hline                                    
 	\end{tabular}}
 	\label{icm_walk}
  \end{table}

	We summarize the RMSEs and MEs of different algorithms under different tests in Table \ref{icmwalkingTest}. One can see that the performances of the correntropy-based algorithms again significantly outperform their traditional counterparts, especially in terms of of MEs.

\begin{figure}[htbp]
		\centerline{\includegraphics[width=8.0cm]{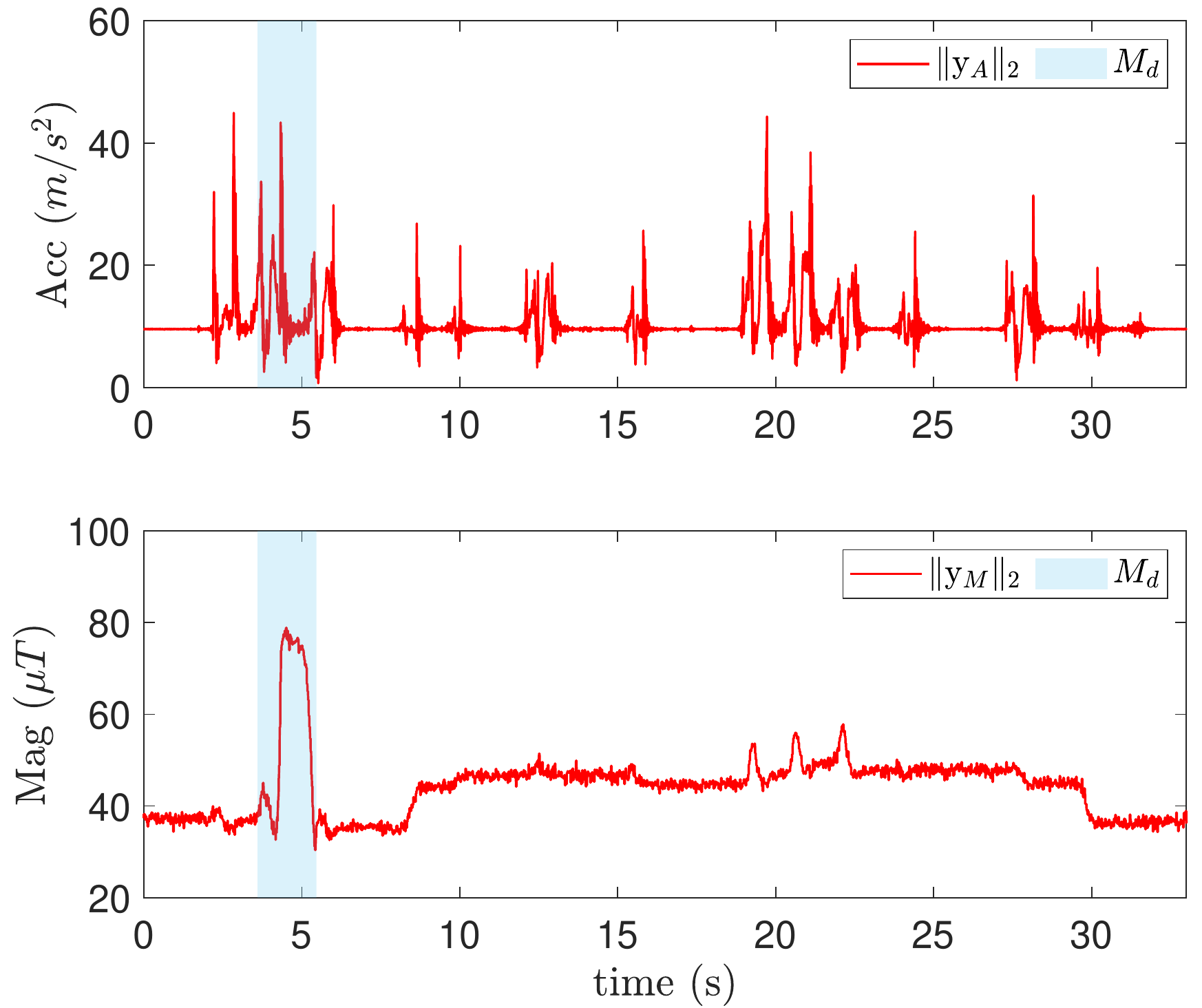}}
		\caption{\textcolor{black}{The norm of sensor readings in Test 4.}}
		\label{icm_lapmag_norm}
	\end{figure}

\begin{figure}[htbp]
 		\centerline{\includegraphics[width=8.0cm]{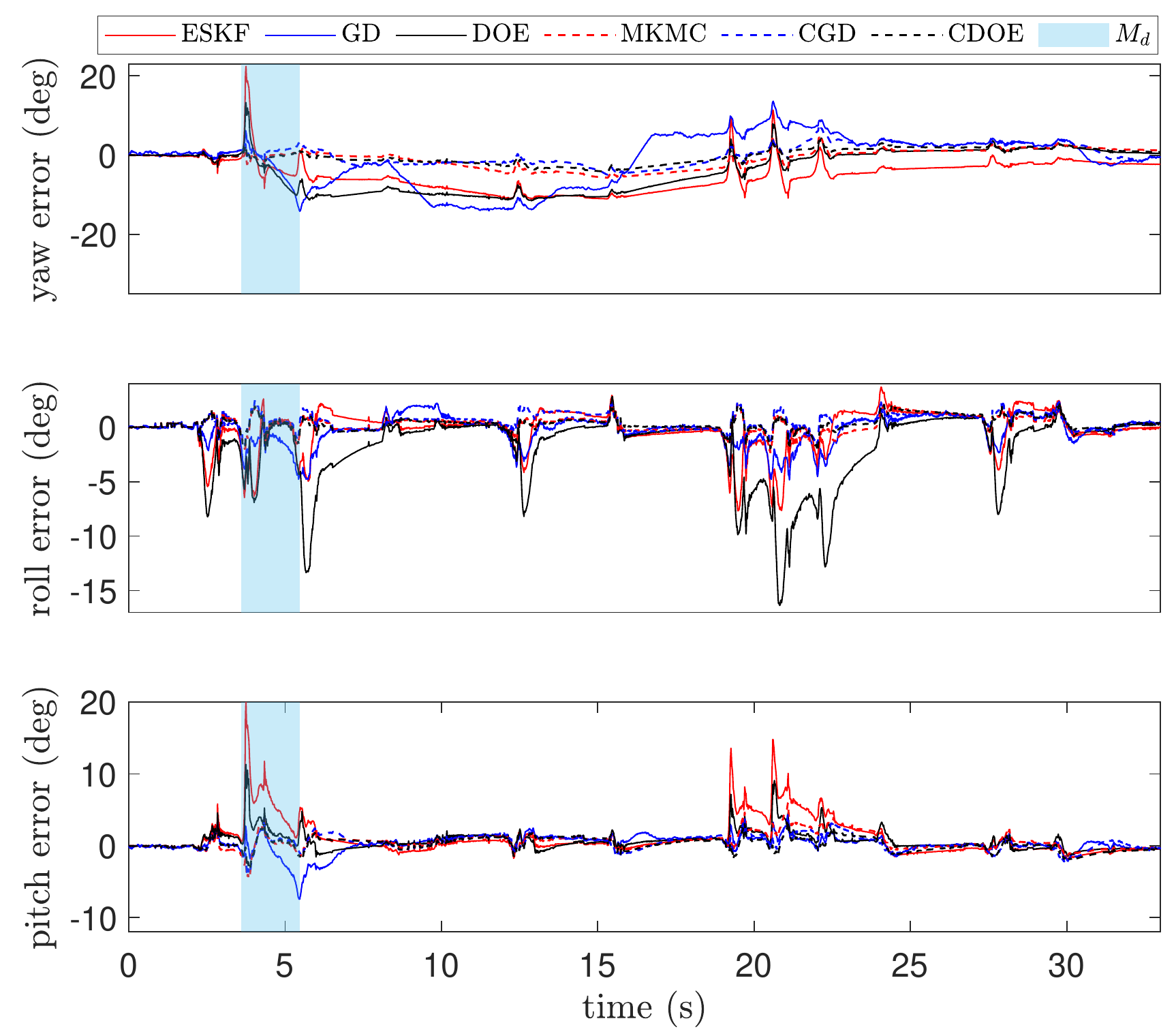}}
 		\caption{The orientation errors of different algorithms in Test 4.}
 		\label{walk_lap_mag}
 	\end{figure}

	We visualize the norm of sensor readings and the error performance of different algorithms under Test 4 in Figs. \ref{icm_lapmag_norm} and \ref{walk_lap_mag}. The magnetic disturbance area is shown in the blue regions. One can see that obvious yaw errors are caused by the ESKF, GD, and DOE in this region, but are avoided by the MKMC, CGD, and CDOE.  

 \subsection{Algorithm Complexity}
 Due to the need of matrix inversion in the ESKF and both matrix inversion and Cholesky decomposition in the MKMC, the algorithm complexity of the ESKF and MKMC is heavier than  $O(n^3)$ where $n=12$ (one can refer to ~\cite{b16} for the detailed algorithm of the MKMC). These time-consuming calculations can be avoided by the CGD and CDOE. To explicitly analyze the time consumption of different algorithms in an embedded low-cost microprocessor, we execute them on a low-cost IMU (computation chip: STM32F405RGT6). The average time consumption in one iteration is obtained by averaging the time cost of 10000 execution iterations and the corresponding results are summarized in Table \ref{ExecutionTime1}. One can see that the complexity of the CGD and CDOE are slightly heavier than the GD and DOE, but are significantly lighter than the MKMC and ESKF. The time costs are as large as 2.515 ms and 3.536 ms for the ESKF and MKMK, but merely 0.059 ms and 0.080 ms for the CGD and CDOE. 	
\begin{table}[htbp]
	\centering
	\caption{Execution Time of Different Methods on a Low-cost IMU.}
	\scalebox{0.85}{
		\begin{tabular}{llllllll}
			\hline
			\hline                                                             Algorithm& {GD} & {CGD} & {DOE} &{CDOE} &{ESKF} & {MKMC} \\
			\hline
			Time (ms)&0.053  & 0.059  & 0.075 & 0.080  & 2.515 &3.536 \\
			\hline
			\hline
	\end{tabular}}
	\label{ExecutionTime1}
\end{table}
 \color{black}
\subsection{Discussion}
	From Tables \ref{linearRota} and \ref{icmwalkingTest}, we observe that the correntropy-based algorithms (i.e., the CGD, CDOE, and MKMC) are significantly better than their counterparts (i.e., the GD, DOE, and ESKF). The effectiveness of the proposed algorithms implies that many existing orientation algorithms can be further improved by re-modeling the objective functions so that the cost-induced distribution much more matches the underlying noise distribution. ``Optimizing" the objective function is powerful but is not fully explored in the literature, especially in the field of sensor fusions where the LS criterion is widely used. Our work demonstrates that the MKCL is a suitable candidate when the underlying noise is heavy-tailed. It is worth mentioning that the Gaussian kernel is not the only kernel function for the formulation of the MKCL, but also some other kernels (e.g., generalized Gaussian kernels, Cauchy kernels). Compared with the loss functions utilized in robust statistics, e.g., the Huber loss, least absolute loss, and elastic net, the MKCL is much more flexible (by varying kernel bandwidths) and possesses a redescending influence function. Other types of non-convex penalties should also be useful as long as they can induce suitable heavy-tailed distributions, which gives great freedom to the designers.

	We observe that the performance of the CDOE is slightly better than the CGD in our experiments. One reason may be that the modeling of the gyroscope bias is considered in the CDOE but ignored in the CGD. Another reason may be that although the proposed methods are very effective in rejecting big disturbances, they are less effective in handling small but lasting disturbances. In many indoor applications, the magnetic field is not strictly homogeneous and may vary with position (one can see the norm of the magnetometer readings in Fig. \ref{icm_lapmag_norm}). In this case, the CDOE outperforms the CGD since it constrains that the magnetic readings only affect the heading, but the small lasting magnetic disturbance may deteriorate both the pitch and roll in the CDOE. Conceptually, it is hard to distinguish the small disturbance from the nominal noise since they share similar regions in the density functions. One possible solution may be extending the zero-mean Gaussian kernel to the variable center Gaussian kernel so that the small disturbance can be mitigated by the shifted centers, which may be our future work. 
	  \color{black}
\section{Conclusion}
	\color{black}
	In this paper, we build a connection between the MKCL and its induced noise distribution and demonstrate that this distribution becomes Gaussian with infinite kernel bandwidth. Some important properties of the MKCL as a cost function are given. Moreover, two MKCL-based algorithms (i.e., the CGD and CDOE) are derived for the orientation estimation of IMUs. The proposed approaches exhibit robustness to external disturbances, and their performances are verified under extensive experiments. Our proposed two methods bear significantly less complexity compared with the ESKF and MKMC, which should be beneficial when the microprocessor computation resource is limited and can save power consumption in a practical implementation. To avoid the complicated kernel bandwidth procedure, in the future, we will focus on the adaptive kernel bandwidth tuning strategies. Additionally, we plan to extend the Gaussian kernel to the variable center Gaussian kernel to reject the small but lasting disturbances.
	\color{black}
  \section{Appendix}
   \subsection{Proof of Theorem \ref{theorem1}}
  \label{proof1}
	\begin{proof}
	Taking Taylor series expansion of $G_{\sigma_i}\big(e_{i,k}\big)$ gives
	\begin{equation}\nonumber
		G_{\sigma_i}\big(e_{i,k}\big)=\sum_{n=0}^{\infty}\frac{(-1)^{n}}{2^{n}{\sigma}_{i}^{2n}n!}e_{i,k}^{2 n}
	\end{equation}
	By setting $\sigma_i \to \infty$, it follows 
	\begin{equation}
		\lim\limits_{\sigma_i \to \infty} \sigma_i^2 \Big(1-G_{\sigma_i}\big(e_{i,k}\big)\Big) = e_{i,k}^2/2
	\end{equation}
	Substituting this result into \eqref{GL}, one obtains 
	\begin{equation}
		\lim\limits_{\sigma_i \to \infty} J_{GL}= \frac{1}{N}\sum_{k=1}^{N}\sum_{i=1}^{l}e_{i,k}^2/2=J_{LS}.
	\end{equation} 
	This completes the proof.
  \end{proof}
  \subsection{Proof of Theorem \ref{theorem2}}
  \label{proof2}
  \begin{proof}
  	If $\tilde{e}_k \sim \mathcal{N}(0,I)$, one has $p(\tilde{e}_{i,k})=\frac{1}{\sqrt{2\pi}}\exp\big(-\frac{\tilde{e}_{i,k}^2}{2}\big)$. Then, the likelihood of $x$ given the set $\{y_k,u_k\}_{k=1}^{N}$ follows
  	\begin{equation}\nonumber
  	\begin{aligned}
  	\mathcal{L}\Big(x;\{y_k,u_k\}_{k=1}^{N}\Big)&=\prod_{k=1}^{N}p\big(\tilde{e}_k\big)\\
  	&=\prod_{k=1}^{N} \prod_{i=1}^{l}\frac{1}{\sqrt{2\pi}}\exp\Big(-\frac{\tilde{e}_{i,k}^2}{2}\Big).
  	\end{aligned}
  	\end{equation}
  	Based on MLE, it follows that 
  	$$
  	x = \arg \max\limits_{x} \mathcal{L}\Big(x;\{y_k,u_k\}_{k=1}^{N}\Big)
  	$$
  	which is equivalent to minimizing its negative logarithm function (ignoring
  	the normalization constants) with
  	$$
  	x = \arg \min\limits_{x} \frac{1}{2}\sum_{k=1}^{N}\sum_{i=1}^{l}\tilde{e}_{i,k}^2.
  	$$
  	Multiplying a constant $1/N$ on the right side of the above equation, we obtain \eqref{mseobj}. This reveals that LS is an optimal metric when the noise is Gaussian under the MLE. On the contrary, if $\tilde{e}_{i,k}$ is heavy-tailed and its density function $p(\tilde{e}_{i,k})$ follows \eqref{pdfe}, one has
  	\begin{equation}
  	\small
  	\begin{aligned}
  	{x} &= \arg \max \limits_{x} \sum_{k=1}^{N}\sum_{i=1}^{l}c_i \exp\Big(-\sigma_i^2(1-\exp\big(-\frac{\tilde{e}_{i,k}^2}{2\sigma_i^2})\big)\Big)\\  	
  	&=\arg \min\limits_{x}\sum_{k=1}^{N}\sum_{i=1}^{l}\sigma_i^2 \Big(1- G_{\sigma_i}(\tilde{e}_{i,k})\Big)\\
  	&=\arg \min\limits_{x}\frac{1}{N}\sum_{k=1}^{N}\sum_{i=1}^{l}\sigma_i^2 \Big(1-G_{\sigma_i}(\tilde{e}_{i,k})\Big)\\
  	&=\arg \min_x J_{CL}(\tilde{e}_k).
  	\label{globj}
  	\end{aligned}
  	\end{equation}
  	This indicates that MKCL is an optimal metric when $\tilde{e}_{i,k}$ follows \eqref{pdfe} and hence completes the proof.
  \end{proof}

\bibliographystyle{IEEEtran}
\bibliography{MCC_GD_TIM_1}

\end{document}